\documentclass[prl,twocolumn,superscriptaddress,longbibliography]{revtex4-2}
\usepackage{amsmath,amssymb,bm}
\usepackage{graphicx}
\usepackage{placeins}
\usepackage{longtable}

\newlength{\SMwidecol}
\newlength{\SMnarrowcol}
\newcommand{\SMmaterialcell}[1]{%
  \parbox[t]{\SMwidecol}{\raggedright\strut #1\strut\par}%
}
\newcommand{\SMidcell}[1]{%
  \parbox[t]{\SMwidecol}{\raggedright\strut #1\strut\par}%
}
\usepackage[colorlinks=true,linkcolor=blue,citecolor=blue,urlcolor=blue]{hyperref}
\usepackage{bibunits}
\newcommand{\PT}{\mathcal{PT}}
\newcommand{\T}{\mathcal{T}}
\newcommand{\vv}[1]{\bm{#1}}
\newcommand{\kx}{k_x}
\newcommand{\ky}{k_y}
\newcommand{\ee}[1]{e^{#1}}

\newif\ifSupplementTOC
\AtBeginDocument{%
  \let\OriginalAddToContents\addtocontents
  \renewcommand{\addtocontents}[2]{%
    \def\RequestedContentsFile{#1}%
    \def\MainContentsFile{toc}%
    \ifx\RequestedContentsFile\MainContentsFile
      \ifSupplementTOC
        \OriginalAddToContents{#1}{#2}%
      \fi
    \else
      \OriginalAddToContents{#1}{#2}%
    \fi
  }%
}

\begin{document}

\title{Néel-Vector Control of the Josephson Diode Effect in
$\mathcal{PT}$-symmetric Antiferromagnets}

\author{Xian-Tang Xu}
\altaffiliation{These authors contributed equally to this work.}
\affiliation{Anhui Province Key Laboratory of Low-Energy Quantum Materials and Devices, High Magnetic Field Laboratory, HFIPS, Chinese Academy of Sciences, Hefei, Anhui 230031, China}
\affiliation{Science Island Branch of Graduate School, University of Science and Technology of China, Hefei, Anhui 230026, China}

\author{Xun-Jiang Luo}
\email{xjluo@hmfl.ac.cn}
\altaffiliation{These authors contributed equally to this work.}
\affiliation{Anhui Province Key Laboratory of Low-Energy Quantum Materials and Devices, High Magnetic Field Laboratory, HFIPS, Chinese Academy of Sciences, Hefei, Anhui 230031, China}

\author{Mingliang Tian}
\affiliation{Anhui Province Key Laboratory of Low-Energy Quantum Materials and Devices, High Magnetic Field Laboratory, HFIPS, Chinese Academy of Sciences, Hefei, Anhui 230031, China}

\author{Ning Hao}
\email{haon@hmfl.ac.cn}
\affiliation{Anhui Province Key Laboratory of Low-Energy Quantum Materials and Devices, High Magnetic Field Laboratory, HFIPS, Chinese Academy of Sciences, Hefei, Anhui 230031, China}

\begin{abstract}
Although $\mathcal{PT}$ symmetry enforces twofold band degeneracy, it does not preclude momentum-asymmetric dispersion when inversion and time-reversal symmetries are individually broken. Here, we show that this provides a distinct route to field-free Josephson nonreciprocity in junctions formed by conventional $s$-wave superconductors and a $\mathcal{PT}$-symmetric collinear antiferromagnet modeled on CuMnAs. Using microscopic modeling and symmetry analysis, we show that these junctions
exhibit both the Josephson diode effect and $\varphi_{0}$-junction
states controlled by the N\'eel vector:
rotating it from $x$ to $y$ switches both effects off, whereas, reversing it reverses the diode polarity.  To reveal the microscopic mechanism, we
develop a channel-resolved scattering theory that accurately captures
the anomalous phases and establishes the  condition for the diode
effect. The multichannel current--phase relations collectively yield a
sizable diode efficiency, tunable by both the magnitude and direction of
the exchange field. Furthermore, a Green-function reduction identifies a
single renormalized $\PT$-degenerate band as the transport carrier and 
quantitatively accounts for the full current amplitudes. These results establish $\PT$-symmetric antiferromagnets as versatile
platforms for field-free, highly tunable Josephson diodes and $\varphi_{0}$ junctions.
\end{abstract}

\maketitle

	A Josephson junction (JJ) carries a dissipationless supercurrent governed
	by the phase difference $\varphi$ between two weakly coupled
	superconductors~\cite{Josephson1962,Golubov2004}. In conventional junctions that preserve
	either inversion $\mathcal{P}$ or time reversal $\T$, the current--phase
	relation (CPR) is constrained to be odd, $I(\varphi)=-I(-\varphi)$, so
	that the critical currents in the two directions are equal. When both
	symmetries are broken, this constraint is lifted: the forward and backward
	critical currents can differ, $I_c^{+} \neq |I_c^{-}|$, giving rise to the
	Josephson diode
	effect~\cite{Nadeem2023,Davydova2022,Zhang2022General,Wang2025}. The same
	symmetry breaking permits a closely related phenomenon, the $\varphi_0$
	junction, whose ground-state phase is neither $0$ nor $\pi$ and which
	enables phase batteries and offset-free qubit
	elements~\cite{Buzdin2008,Szombati2016,Assouline2019,Strambini2020}. Both effects typically arise from
	the interplay of magnetism and spin--orbit coupling~\cite{Daido2022,YuanFu2022,He2022} and have been realized
	in diverse platforms, including Rashba spin--orbit-coupled
	systems~\cite{Baumgartner2022,CostaFabian2023,Lotfizadeh2024,Su2024,Bhowmik2025,Zhu2025Andreev,Mangold2026}, topological
	materials~\cite{Tanaka2022DwaveTI,Pal2022,Lu2023TI,Legg2023Parity,Anh2024,Le2024,Zeng2025TiltedDirac,McFarlane2026}, van der Waals
	heterostructures~\cite{Wu2022,Bauriedl2022,Hu2023,DiezMerida2023,Volkov2024Twisted,Ghosh2024HighT,Ma2025NbSe2,Wang2026TwistedFeSe}, and symmetry-compensated magnets \cite{Ouassou2023,BeenakkerVakhtel2023,Lu2024,Cheng2024,Chakraborty2025,LiHuZhang2025,Yang2025NiI2,ChengHelimagnet2026}. Unlike altermagnetic Josephson diodes, where \(\mathcal{PT}\) symmetry is broken and spin-split bands play a central role, the present mechanism operates in a \(\mathcal{PT}\)-preserving system with twofold-degenerate bands.This nonreciprocal
	supercurrent enables rectification without dissipation and holds promise
	for superconducting electronics~\cite{Nadeem2023,Lou2026Kagome}.

	$\PT$-symmetric antiferromagnets represent a broad class of magnetic
	systems in which both $\mathcal{P}$ and $\T$ are individually broken
	while their product $\PT$ is retained~\cite{Baltz2018}. Although $\PT$
	enforces a twofold degeneracy at every momentum, the individual
	breaking of $\mathcal{P}$ and $\T$ allows the dispersion to become
	asymmetric, $\varepsilon(\vv{k}) \neq \varepsilon(-\vv{k})$~\cite{Iwata2026}.
	The same symmetry breaking gives rise to a variety of nonreciprocal
	phenomena, ranging from charge
	transport~\cite{Tokura2018,PhysRevLett.129.276601,Ye2022} and acoustic
	phonons~\cite{Ren2025} to the nonlinear Hall
effect~\cite{PhysRevLett.127.277201,PhysRevLett.127.277202,doi:10.1126/science.adf1506,Wang2023}.
	These observations naturally raise an open question: can the same
	symmetry breaking---and the asymmetric band structure it
	produces---give rise to a superconducting diode effect  in
	$\PT$-symmetric antiferromagnets? More importantly, can the electrical
	tunability of the N\'eel vector in these
	systems~\cite{Wadley2016,Godinho2018} serve as a knob to tune the superconducting diode effect,
	a capability essential for superconducting device applications?

\begin{figure*}[t]
\centering
\includegraphics[width=\textwidth]{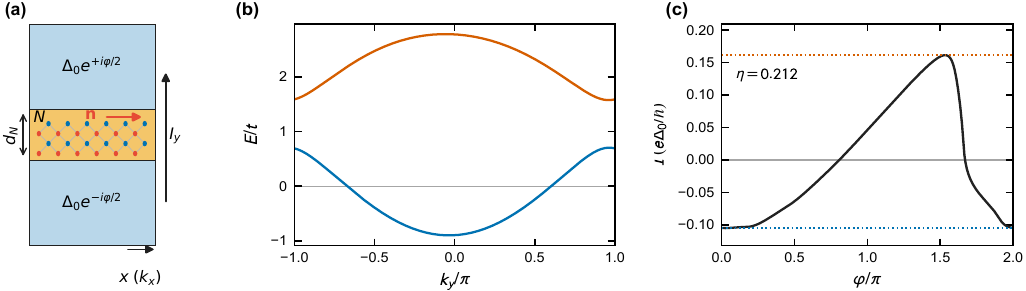}
\caption{(a) Two-sublattice antiferromagnetic SNS junction. Red $A$ sites carry
$+J_n\vv{n}$ and blue $B$ sites carry $-J_n\vv{n}$. (b) Full normal-state spectrum \(E_{\zeta\mathbf{k}}\) for the two branches \(\zeta=\pm1\) as a function of \(k_y\) at fixed \(k_x=0.3\pi\) and \(\boldsymbol n\parallel\hat{x}\). Energies are in units of \(t\). (c) Complete finite-SNS CPR at $\mu=-1.1t$ and $J_n=0.4t$.
The unequal positive and negative critical-current magnitudes
give $\eta=0.212$.
The common parameters are $t^{\prime}=0.1t$,
$\Delta_0=0.075t$, $\lambda_N=0.25t$, $d_N=12$, and $W_S=20$.
The superconducting slabs have $J_S=\lambda_S=0$ and the NS interface bonds have zero SOC.The parameters used in all figures are described in the SM.}
\label{fig:1}
\end{figure*}

In this Letter, we study Josephson junctions composed of conventional
$s$-wave superconductors and a $\PT$-symmetric collinear antiferromagnet
modeled on CuMnAs
[Fig.~\ref{fig:1}(a)]. The diode effect and the $\varphi_{0}$-junction
state emerge together once the transport-reversing magnetic mirror
$\mathcal{M}_y$ is broken. Specifically, for a N\'eel vector $\vv{n}\parallel\hat{x}$, $\mathcal{M}_y$
is broken and the $\PT$-degenerate bands become asymmetric along the
transport direction [Fig.~\ref{fig:1}(b)], producing anomalous phase
shifts and unequal critical currents [Fig.~\ref{fig:1}(c)]. For
$\vv{n}\parallel\hat{y}$, by contrast, $\mathcal{M}_y$ is preserved and
neither effect can occur. Consequently, the N\'eel vector acts as a
control knob: rotating it by $90^{\circ}$ toggles both effects on and
off, whereas reversing it flips the diode polarity. Conversely, the
diode effect provides a readout of the N\'eel vector.  To uncover the microscopic mechanism, we develop an analytic,
channel-resolved scattering theory on the Matsubara axis to reveal the phase structure of the CPR. The theory accurately captures
the anomalous phases and establishes the condition for the diode
effect. Although the individual channels are weak diodes, their
coherent sum yields a sizable efficiency whose sign and magnitude track
the direction and strength of the exchange field. We
further employ a Green-function reduction to recover the current
amplitudes, tracing the entire transport to a single renormalized
$\PT$-degenerate band. Our work establishes $\PT$-symmetric
antiferromagnets as versatile platforms for field-free Josephson diodes and
$\varphi_{0}$ junctions, and demonstrates that electrical control of the
N\'eel vector provides a practical route to manipulating both effects.

	\emph{Model and asymmetric bands.---}We consider the two-sublattice
	antiferromagnetic SNS junction, as schematically illustrated in Fig.~\ref{fig:1}(a). Two conventional
	$s$-wave superconducting slabs of width $W_S$ are separated by a weak
	link of width $d_N$, a strip of the collinear
	antiferromagnet modeled on CuMnAs. The interfaces are parallel to $x$ and
	the current flows along $y$. In the Nambu basis
	$\Psi_{\vv{k}}=(\psi_{A,\vv{k}},\psi_{B,\vv{k}})^{T}$,
	$\psi_{\alpha,\vv{k}}=(c_{\alpha\vv{k}u},c_{\alpha\vv{k}d},
	c^{\dagger}_{\alpha,-\vv{k}d},-c^{\dagger}_{\alpha,-\vv{k}u})^{T}$,
	where $\alpha=A,B$ labels the sublattices and $u,d$ label spin up and
	down, the Bogoliubov--de Gennes (BdG) Hamiltonian
	reads~\cite{Smejkal2017,Hu2026PTJosephson}
	\begin{align}
	H_{\mathrm{BdG}}&=\big[\xi_{\vv{k}}\rho_{0}+q_{\vv{k}}\rho_{x}\big]\tau_{z}
	+\lambda(y)\rho_{z}\tau_{z}\big(\sin\kx\,\sigma_{y}-\sin\ky\,\sigma_{x}\big)
	\nonumber\\
	&\quad+J_{n}(y)\rho_{z}\,\vv{n}\cdot\vv{\sigma}
	+\Delta(y)\tau_{+}+\Delta^{*}(y)\tau_{-}.
	\label{eq:1}
	\end{align}
	Here $\rho_i$, $\sigma_i$, and $\tau_i$ act in sublattice, spin, and
	particle--hole spaces, respectively, and
	$\tau_{\pm}=(\tau_x\pm i\tau_y)/2$. The intersublattice and
	intrasublattice hoppings $t$ and $t'$ give
	$\xi_{\vv{k}}=-\mu-t'(\cos\kx+\cos\ky)$ and
	$q_{\vv{k}}=-2t\cos(\kx/2)\cos(\ky/2)$, and the N\'eel vector lies in
	the plane, $\vv{n}=(\cos\theta,\sin\theta,0)$. With the junction
	centered at $y=0$, all spatial profiles are fixed by
	\begin{align}
	&\chi_{N}(y)=\Theta(d_{N}/2-|y|),\nonumber\\
	&J_{n}(y)=J_{n}\chi_{N}(y),\qquad\lambda(y)=\lambda\chi_{N}(y)\nonumber\\
	&\chi_{L}(y)=\Theta(-y-d_{N}/2)\,\Theta(y+d_{N}/2+W_{S}),\nonumber\\
	&\chi_{R}(y)=\Theta(y-d_{N}/2)\,\Theta(d_{N}/2+W_{S}-y),\nonumber\\
	&\Delta(y)=\Delta_{0}\left[\ee{-i\varphi/2}\chi_{L}(y)
	+\ee{i\varphi/2}\chi_{R}(y)\right],
	\label{eq:2}
	\end{align}
	where $J_n$ is the staggered exchange strength and $\Delta_0$ is the
	pairing amplitude of the two superconductors with phase difference
	$\varphi$.

	The combined operation $\PT=\rho_x(i\sigma_y)\mathcal{K}$ leaves the
	momentum $\vv{k}$ invariant and obeys $(\PT)^2=-1$, enforcing a
	twofold degeneracy at every momentum. The normal state of the weak
	link hosts two branches, each twofold degenerate,
	\begin{align}
&E_{\zeta\vv{k}}=\xi_{\vv{k}}+\zeta D_{\vv{k}},\quad \zeta=\pm1,
\nonumber\\
&D_{\vv{k}}^{2}=q_{\vv{k}}^{2}+J_{n}^{2}
+\lambda^{2}\!\left(\sin^{2}\!\kx+\sin^{2}\!\ky\right)\nonumber\\
&\qquad\quad+2\lambda J_{n}\!\left(\sin\kx\,\sin\theta-\sin\ky\,\cos\theta\right).
\label{eq:3}
\end{align}
	The lower branch $\zeta=-1$, the bonding-like combination of the two
	sublattices, is the low-energy band that carries the Josephson
	current. Its asymmetry along the transport direction is quantified by
	\begin{align}
	\Delta E_{\zeta}(\kx,\ky)=E_{\zeta}(\kx,\ky)
	-E_{\zeta}(\kx,-\ky),\nonumber\\
    =-\zeta\,\frac{4\lambda J_{n}\cos\theta\,\sin\ky}
	{\sqrt{D^{2}(\kx,\ky)}+\sqrt{D^{2}(\kx,-\ky)}}
	\label{eq:4}
	\end{align}
	Thus, the band asymmetry is controlled entirely by the N\'eel orientation:  $\Delta E_{\zeta}=0$ 
	 at $\theta=\pi/2$ ($\vv{n}\parallel\hat{y}$), whereas it is finite for
	$\vv{n}\parallel\hat{x}$ [Fig.~\ref{fig:1}(b)]. This orientation dependence is dictated by the magnetic mirrors: the
transport-reversing mirror $\mathcal{M}_y=i\sigma_yR_y$ is preserved for
$\vv{n}\parallel\hat{y}$ but broken for $\vv{n}\parallel\hat{x}$, where $R_{y}$ flips $y$.
Therefore, an asymmetric band along the
junction requires the breaking of $\mathcal{M}_y$ as well as $\mathcal{P}$ and $\mathcal{T}$.

\begin{table}[t]
\caption{Symmetry status and resulting anomalous response of the
$y$-directed junction for the two high-symmetry N\'eel orientations.
$\T$, $\mathcal{P}$, and $\mathcal{M}_y$ reverse the transport current;
$\PT$ and $\mathcal{M}_x$ leave it invariant and impose no constraint. The
surviving mirror $\mathcal{M}_y$ for $\vv{n}\parallel\hat{y}$ forbids
both the $\varphi_0$ state and the diode effect.}
\label{tab:1}
\begin{ruledtabular}
\begin{tabular}{lccc}
Operation & Reverses $I_y$? & $\vv{n}\parallel\hat{x}$ & $\vv{n}\parallel\hat{y}$\\
\hline
$\T$ & yes & broken & broken\\
$\mathcal{P}$ & yes & broken & broken\\
$\mathcal{M}_y$ & yes & broken & preserved\\
$\PT$ & no & preserved & preserved\\
\hline
$\varphi_0$ junction &  & allowed & forbidden\\
Diode effect &  & allowed & forbidden\\
\end{tabular}
\end{ruledtabular}
\end{table}

\begin{figure*}[t]
\centering
\includegraphics[width=\textwidth]{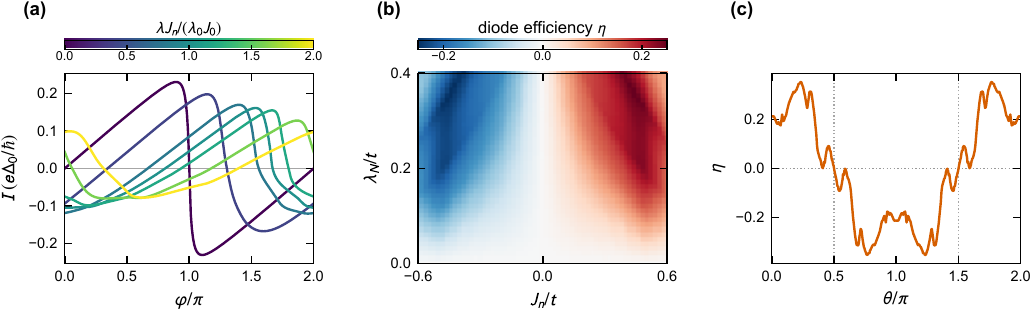}
\caption{(a) Full CPR for increasing $\lambda_N J_n/(\lambda_0J_0)=0,0.4,0.8,1,1.2,1.6,2$ at fixed
$J_n/\lambda_N=1.6$. (b) Total diode efficiency $\eta$ in the $J_n$--$\lambda_N$ parameter plane. (c) Diode efficiency under a full in-plane
rotation of the N\'eel vector at $J_n=0.4t$ and $\lambda_N=0.25t$. The dotted
lines mark the mirror-symmetric orientations $\theta=\pi/2$ and $3\pi/2$,
where the response vanishes.}
\label{fig:control}
\end{figure*}

\emph{Tunable $\varphi_0$ junction and nonreciprocal current.---}The
band nonreciprocity controlled by the N\'eel vector provides the microscopic ingredient for the
diode effect and the $\varphi_0$-junction states. The link is the mirror
$\mathcal{M}_y$: it reverses the transport current and maps
$\varphi\rightarrow-\varphi$. Therefore, $\mathcal{M}_y$ yields the constraint
$I(\varphi)=-I(-\varphi)$, which enforces $I_c^{+}=|I_c^{-}|$ and pins
the equilibrium phase at a pair of additive inverses. Here,
$I_c^{+}$ and $I_c^{-}$ are the maximal positive and negative supercurrents, respectively. Consequently, both the diode effect and
	the $\varphi_{0}$-junction state are forbidden for
	$\vv{n}\parallel\hat{y}$, where $\mathcal{M}_y$ is preserved, and
	allowed for $\vv{n}\parallel\hat{x}$, where $\mathcal{M}_y$ is broken;
	Table~\ref{tab:1} summarizes these symmetry constraints.
	Figure~\ref{fig:1}(c) confirms this analysis numerically: for
	$\vv{n}\parallel\hat{x}$, the full BdG CPR exhibits both effects at
	once---its stable zero shifts away from $0$ and $\pi$, and its extrema
	are unequal, $I_c^{+}\neq|I_c^{-}|$---whereas for
	$\vv{n}\parallel\hat{y}$ the CPR remains odd and neither feature
	appears, as detailed in the Supplemental Material (SM) \cite{SM}. Thus, the asymmetric band and the superconducting diode effect
	share the same symmetry origin.

To quantify both effects, we define the anomalous phase $\varphi_{0}$
as the stable zero of the CPR with positive slope, and the diode
efficiency
\begin{equation}
\eta=\frac{I_c^{+}-|I_c^{-}|}{I_c^{+}+|I_c^{-}|}.
\label{eq:eta}
\end{equation}
Both $\varphi_{0}$ and $\eta$ are continuously tunable.
Figure~\ref{fig:control}(a) shows the numerical CPRs as $\lambda J_n$ increases: the anomalous phase
$\varphi_{0}$ moves continuously around the phase circle, and
$\partial I/\partial\varphi$ at $\varphi=0^{+}$ changes sign. Thus, the
junction undergoes a generalized $0$--$\pi$ transition as $\lambda J_n$
increases. The same control reshapes the diode response
[Fig.~\ref{fig:control}(b)]: the sign of $\lambda J_n$ sets the diode
polarity, whereas $|\eta|$ varies nonmonotonically with its magnitude.  In the SM~\cite{SM}, we further show that a vertical electric field in the weak-link region can tune the diode efficiency; the induced sublattice potential $-v\rho_z\tau_z$ explicitly breaks $\PT$ for $v\neq0$.
Rotating the N\'eel vector provides a direct symmetry test. At
fixed $\lambda$ and $J_n$, $\eta$ varies nonmonotonically with the
N\'eel-vector angle $\theta$ [Fig.~\ref{fig:control}(c)], but symmetry
pins
\begin{equation}
\eta(\pi/2)=\eta(3\pi/2)=0,\qquad \eta(\theta+\pi)=-\eta(\theta).
\label{eq:13}
\end{equation}
Both relations follow directly from the symmetry analysis above: a
$90^{\circ}$ rotation from $\vv{n}\parallel\hat{x}$ restores
$\mathcal{M}_y$ and switches the diode off, while a $180^{\circ}$
reversal flips its polarity.

	\emph{Scattering theory.---}To reveal the microscopic origin of the
	diode effect and the $\varphi_{0}$ junction, we develop a
	channel-resolved scattering theory on the Matsubara axis~\cite{Hu2023}. At fixed
	$\kx$, let $k_{R}(E;\kx)$ and $k_{L}(E;\kx)$ denote the right- and
	left-moving wave vectors on the low-energy band of Eq.~\eqref{eq:3} at
	energy $E$. In the finite lattice junction, the reference planes of the
	first and last normal layers are separated by
	$L_N=(d_N-1)a$, where $a$ is the layer spacing. With momenta measured in
	units of $a^{-1}$, the electron--hole loop then accumulates the
	propagation phase (see the SM \cite{SM})
\begin{equation}
\Theta_{\kx}(E)=L_{N}\left[k_{R}(E;\kx)+k_{L}(-E;\kx)\right],
\label{eq:7}
\end{equation}
and the phase accumulated through a complete loop, closed by two
transparent Andreev reflections at the interfaces, is
\begin{equation}
\vartheta_{\kx}(E)=\varphi+\Theta_{\kx}(E)-2\arccos(E/\Delta_0),
\label{eq:8}
\end{equation}
 We continue the loop phase analytically to the Matsubara axis,
	$E\rightarrow i\omega_n$ with $\omega_n=(2n+1)\pi/\beta$ the fermionic
	Matsubara frequency, and write the continued propagation phase as
	$\Theta_{\kx}(i\omega_n)=\alpha_{\kx}(\omega_n)+i\beta_{\kx}(\omega_n)$.
	Each round trip then factorizes into a modulus and a phase,
	\begin{equation}
	\Gamma_{\kx}(\omega_n)=\ee{i\vartheta_{\kx}(i\omega_n)}
	\equiv-\rho_{\kx}(\omega_n)\,\ee{i[\varphi+\alpha_{\kx}(\omega_n)]},
	\label{eq:g}
	\end{equation}
	  and the Andreev-reflection and propagation decays
	combine into the round-trip attenuation
	\begin{equation}
	\rho_{\kx}(\omega_n)=\exp\left[-2\,\mathrm{arsinh}
	\left(\frac{\omega_n}{\Delta_0}\right)-\beta_{\kx}(\omega_n)\right].
	\label{eq:rho}
	\end{equation}
With this in hand, we can derive the current carried by one $\kx$ channel
by summing the contributions of the two partner loops over Matsubara frequencies~\cite{SM},
	\begin{align}
	I^{\mathrm{M}}_{\kx}(\varphi)&=\frac{4eg_{\kx}}{\hbar\beta}
	\sum_{\omega_n>0}\rho_{\kx}(\omega_n)
	\sin[\varphi+\alpha_{\kx}(\omega_n)]\nonumber\\
	&\times\left\{1+2\rho_{\kx}(\omega_n)
	\cos[\varphi+\alpha_{\kx}(\omega_n)]
	+\rho_{\kx}^{2}(\omega_n)\right\}^{-1},
	\label{eq:6}
	\end{align}
	where $g_{k_x}=2$ accounts for the twofold band degeneracy in the $\PT$-symmetric case $v=0$.
The real part $\alpha_{k_x}(\omega_n)$ sets the phase shift of
each Matsubara contribution, whereas $\rho_{k_x}(\omega_n)$
controls its weight and harmonic content.

\begin{figure}[t]
\centering
\includegraphics[width=\columnwidth]{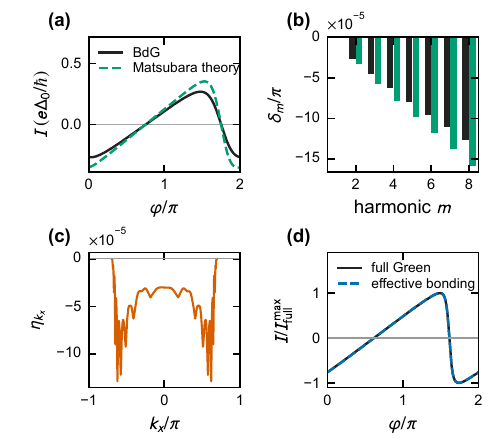}
\caption{(a) Fixed-channel CPR from BdG and the low-energy Matsubara theory
at $\kx=0.3\pi$ and $k_{B}T=0.05\Delta_0$. (b) Corresponding harmonic
mismatch (black and green as in panel (a)),
$\delta_m=\vartheta_m-m\vartheta_1$. (c)
Channel-resolved diode efficiency $\eta_{\kx}$ at $T=0$; channels with current below $10^{-6}$ of the largest channel scale are omitted.
(d) Fixed-channel CPR at $\kx=0$ from the full Green function and the
effective bonding theory obtained through the Schur complement at $T=0$. The remaining parameters are the same as in Fig.~\ref{fig:1}. }
\label{fig:mechanism}
\end{figure}

	The harmonic content of Eq.~\eqref{eq:6} follows from its Fourier
	series,
	\begin{align}
	&I_{\kx}(\varphi)=\mathrm{Im}\sum_{m\geq1}\ee{im\varphi}\mathcal{C}_{m,\kx},
	\nonumber\\
	&\mathcal{C}_{m,\kx}\propto(-1)^{m+1}\sum_{\omega_n>0}
	\rho_{\kx}^{m}(\omega_n)\,\ee{im\alpha_{\kx}(\omega_n)},
	\label{eq:10}
	\end{align}
	so the $m$th harmonic samples the loop phase $\alpha_{\kx}(\omega_n)$
	with its own weight $\rho_{\kx}^{m}(\omega_n)$, which decays
	increasingly rapidly with $\omega_n$ as $m$ grows. Because the loop
	phase itself varies with $\omega_n$, the harmonics need not be phase
	locked. Their mismatch
	\begin{equation}
	\delta_{m,\kx}=\arg\mathcal{C}_{m,\kx}-m\arg\mathcal{C}_{1,\kx}
	\pmod{\pi}
	\label{eq:10b}
	\end{equation}
	diagnoses the channel diode effect: Harmonic phase unlocking prevents the CPR from being made odd by
a shift of the superconducting phase. In the absence of additional
current-reversing symmetries or accidental cancellations, it
generically produces a finite Josephson diode effect. For a CPR
containing only the first and second harmonics, with both amplitudes
nonzero, phase unlocking is an exact necessary and sufficient
condition for critical-current nonreciprocity,
$I_c^{+}\neq|I_c^{-}|$. In the two-harmonic approximation,
$\eta_{k_x}\simeq-(A_{2,k_x}/A_{1,k_x})
\sin\delta_{2,k_x}$.
For the representative $k_x=0.3\pi$ channel,
$A_{2,k_x}/A_{1,k_x}=0.336$, whereas
$|\delta_{2,k_x}|=8.41\times10^{-5}$.
Thus, the very small $|\eta_{k_x}|=2.77\times10^{-5}$
originates primarily from the nearly locked harmonic phases,
rather than from weak higher harmonics.
Higher harmonics modify its sign and precise value.

To benchmark the scattering theory, we compute $I_{\kx}(\varphi)$
from Eq.~\eqref{eq:10} and compare it with the full BdG solution for
a representative channel [Fig.~\ref{fig:mechanism}(a)]. The Matsubara
theory almost reproduces the anomalous phase $\varphi_{0}$, yet
substantially overestimates the current amplitude. The Fourier
diagnostic of Eq.~\eqref{eq:10b} likewise reproduces the BdG phase
mismatches $\delta_{m,\kx}$ through eighth order
[Fig.~\ref{fig:mechanism}(b)]. Thus, the scattering theory captures
the phase structure of the CPR but overestimates its magnitude,
because it assumes transparent Andreev reflections at the interfaces.

 The full model further shows that each channel is individually a very
weak diode, with $\max_{\kx}|\eta_{\kx}|=1.29\times10^{-4}$
[Fig.~\ref{fig:mechanism}(c)]. However, the coherent sum over channels,
$I(\varphi)=\sum_{\kx}I_{\kx}(\varphi)$,  yields a sizable
efficiency $|\eta|=0.212$---more than three orders of magnitude above the
single-channel bound. The diode effect is therefore a collective
property of the coherent multichannel junction, not of any single
channel.

	\emph{Green-function reduction.---}To recover the current amplitudes,
	we turn to the exact Green function of the finite SNS junction and map
	the weak link onto a single effective low-energy band. At Matsubara
	frequency $i\omega_n$, the inverse Green function of the normal region
	is $\mathcal{G}_N^{-1}\equiv\mathcal{M}
	=i\omega_n-H_N-\Sigma_S(i\omega_n,\varphi)$ (see the SM \cite{SM}), where the exact
	self-energy $\Sigma_S$ of the superconducting slabs restores the
	finite-superconductor and matrix-interface structure omitted by the
	transparent-interface formula Eq.~\eqref{eq:6}. We partition
	$\mathcal{M}$ into blocks $\mathcal{M}_{ij}=P_i\mathcal{M}P_j$
	($i,j=b,r$): the bonding projector $P_b$ selects the bonding-derived subspace defined by the intersublattice hopping. In this basis,
	\begin{equation}
	\mathcal G_N^{-1}=
	\begin{pmatrix}
	\mathcal M_{bb}&\mathcal M_{br}\\
	\mathcal M_{rb}&\mathcal M_{rr}
	\end{pmatrix},
	\label{eq:main-band-blocks}
	\end{equation}
	and integrating out $r$ exactly yields the Schur complement
	\begin{equation}
	\mathcal{M}^{\mathrm{eff}}_{b}
	=\mathcal{M}_{bb}-\mathcal{M}_{br}\mathcal{M}_{rr}^{-1}\mathcal{M}_{rb},
	\label{eq:11}
	\end{equation}
	whose second term resums all virtual excursions
	$b\rightarrow r\rightarrow b$.
	Figure~\ref{fig:mechanism}(d) validates the channel-resolved reduction: the effective bonding band of Eq.~\eqref{eq:11} exactly reproduces
	the channel-resolved critical currents $I_{c,\kx}$ of the full
	Green-function calculation. The active object is thus a single renormalized
	low-energy band, its frequency-dependent Green function dressed by the
	complementary sector. The full derivation is given in the
	SM~\cite{SM}.

\begin{table}[t]
\caption{Symmetry-allowed band nonreciprocity along the coordinate axes
for all 21 magnetic point groups $G$ containing $\mathcal{PT}$ but neither
$\mathcal P$ nor $\mathcal T$.
$\checkmark$ denotes allowed nonreciprocity and $\times$ denotes
symmetry-enforced reciprocity of the normal-state spectrum.
The coordinate conventions and corresponding magnetic structures are
specified in the final section of the SM.}
\label{tab:pt-directions}
\centering
\begingroup
\setlength{\tabcolsep}{2pt}
\renewcommand{\arraystretch}{1.06}
\begin{tabular*}{\linewidth}{@{\extracolsep{\fill}}cccc@{\hspace{1.2em}}cccc@{}}
\hline\hline
$G$ & $x$ & $y$ & $z$ & $G$ & $x$ & $y$ & $z$ \\
\hline
$\bar{1}'$ & $\checkmark$ & $\checkmark$ & $\checkmark$ & $\bar{3}'m$ & $\times$ & $\checkmark$ & $\checkmark$ \\
$2'/m$ & $\checkmark$ & $\checkmark$ & $\times$ & $\bar{3}'m'$ & $\checkmark$ & $\times$ & $\times$ \\
$2/m'$ & $\times$ & $\times$ & $\checkmark$ & $6/m'$ & $\times$ & $\times$ & $\checkmark$ \\
$m'm'm'$ & $\times$ & $\times$ & $\times$ & $6'/m$ & $\checkmark$ & $\checkmark$ & $\times$ \\
$m'mm$ & $\checkmark$ & $\times$ & $\times$ & $6/m'm'm'$ & $\times$ & $\times$ & $\times$ \\
$4/m'$ & $\times$ & $\times$ & $\checkmark$ & $6/m'mm$ & $\times$ & $\times$ & $\checkmark$ \\
$4'/m'$ & $\times$ & $\times$ & $\times$ & $6'/mmm'$ & $\times$ & $\checkmark$ & $\times$ \\
$4/m'm'm'$ & $\times$ & $\times$ & $\times$ & $m'\bar{3}'$ & $\times$ & $\times$ & $\times$ \\
$4/m'mm$ & $\times$ & $\times$ & $\checkmark$ & $m'\bar{3}'m$ & $\times$ & $\times$ & $\times$ \\
$4'/m'm'm$ & $\times$ & $\times$ & $\times$ & $m'\bar{3}'m'$ & $\times$ & $\times$ & $\times$ \\
$\bar{3}'$ & $\checkmark$ & $\checkmark$ & $\checkmark$ &  &  &  &  \\
\hline\hline
\end{tabular*}
\endgroup
\end{table}

	\emph{Conclusion and discussion.---}In summary, we have established
the $\PT$-symmetric antiferromagnet CuMnAs as a versatile platform for both
the Josephson diode effect and the $\varphi_{0}$-junction state.
Symmetry analysis and microscopic modeling demonstrate full
N\'eel-vector control: a $90^{\circ}$ rotation switches both effects
off, a $180^{\circ}$ rotation flips the diode polarity, and the
efficiency tracks the magnitude and direction of the exchange field.
A channel-resolved scattering theory reveals the anomalous phases and
the  diode condition, while a Green-function reduction identifies
a single renormalized $\PT$-degenerate band as the transport carrier
and reproduces the full current amplitudes.

	We emphasize that room-temperature experiments in CuMnAs have established electrical
	control of the N\'eel vector through a current-induced staggered spin-orbit field \cite{PhysRevLett.113.157201,Wadley2016,Godinho2018}. Specifically, the writing current controls the N\'eel vector at two
	levels: its direction selects the axis of $\vv{n}$, which relaxes to
	the in-plane orientation perpendicular to the current, giving
	reversible $90^{\circ}$ switching between the two orthogonal
	states~\cite{Wadley2016}, whereas its polarity selects the sign of
	$\vv{n}$ along a given axis, giving the $180^{\circ}$
	reversal~\cite{Godinho2018}. The N\'eel-vector control
	proposed here is therefore experimentally feasible: the $90^{\circ}$
	operation toggles the diode effect and the $\varphi_{0}$ junction; the $180^{\circ}$
	operation reverses the diode polarity. Conversely, the diode effect
itself provides an electrical signature readout of the N\'eel vector in an
antiferromagnet---a long-standing challenge given the vanishing net
magnetization.

To extend the symmetry analysis beyond CuMnAs, we classify all 21 magnetic
point groups that preserve $\mathcal{PT}$ while breaking $\mathcal P$ and
$\mathcal T$ separately, and summarize their allowed directions of band
nonreciprocity in Table~\ref{tab:pt-directions}.
Thirteen groups allow nonreciprocity along at least one coordinate axis;
the remaining eight enforce reciprocity along all three axes but can
still permit it along generic oblique directions in three dimensions.
This distinction makes the magnetic domain and junction orientation
essential design choices, since a symmetry that reverses an entire
two-dimensional momentum plane also constrains every oblique channel
within that plane.
In the Supplemental Material~\cite{SM}, we list the corresponding
materials by magnetic point group, using magnetic-structure data
compiled from Supplemental Table XI of Ref.~\cite{Xiao2023MagneticSHG}; this classification
identifies symmetry-compatible band structures, while the Josephson
diode response must additionally be assessed from the symmetries and
current--phase relation of the complete junction.

\begin{acknowledgments}
\emph{Acknowledgments}—This work was supported by the National Key R\&D Program of China (Grant Nos.
2022YFA1403200 and 2024YFA1613200), the National
Natural Science Foundation of China (Grant Nos. 92565201,
92265104, and
12604254), the Basic Research Program
of the Chinese Academy of Sciences Based on Major
Scientific Infrastructures (Grant No. JZHKYPT-
2021-08), the CASHIPS Directors Fund (Grant
No. BJPY2023A09), Anhui Provincial Major S\&T
Project (s202305a12020005), and the High Magnetic
Field Laboratory of Anhui Province under Contract No.
AHHM-FX-2020-02.
\end{acknowledgments}

\bibliographystyle{apsrev4-2}
\bibliography{neel_josephson_diode_references}

@article{Josephson1962,
  author = {Josephson, B.D.},
  title = {{Possible new effects in superconductive tunnelling}},
  journal = {Phys. Lett.},
  volume = {1},
  number = {7},
  pages = {251--253},
  year = {1962},
  doi = {10.1016/0031-9163(62)91369-0},
  url = {https://doi.org/10.1016/0031-9163(62)91369-0}
}

@article{Golubov2004,
  author = {Golubov, A. A. and Kupriyanov, M. Yu. and Il'ichev, E.},
  title = {{The current-phase relation in Josephson junctions}},
  journal = {Rev. Mod. Phys.},
  volume = {76},
  pages = {411--469},
  year = {2004},
  doi = {10.1103/RevModPhys.76.411},
  url = {https://doi.org/10.1103/RevModPhys.76.411}
}

@Article{Tokura2018,
author={Tokura, Yoshinori
and Nagaosa, Naoto},
title={Nonreciprocal responses from non-centrosymmetric quantum materials},
journal={Nature Communications},
year={2018},
month={Sep},
day={14},
volume={9},
number={1},
pages={3740},
issn={2041-1723},
doi={10.1038/s41467-018-05759-4},
url={https://doi.org/10.1038/s41467-018-05759-4}
}

@article{PhysRevLett.127.277201,
  title = {Intrinsic Nonlinear Hall Effect in Antiferromagnetic Tetragonal CuMnAs},
  author = {Wang, Chong and Gao, Yang and Xiao, Di},
  journal = {Phys. Rev. Lett.},
  volume = {127},
  issue = {27},
  pages = {277201},
  numpages = {6},
  year = {2021},
  month = {Dec},
  publisher = {American Physical Society},
  doi = {10.1103/PhysRevLett.127.277201},
  url = {https://link.aps.org/doi/10.1103/PhysRevLett.127.277201}
}

@article{PhysRevLett.127.277202,
  title = {Intrinsic Second-Order Anomalous Hall Effect and Its Application in Compensated Antiferromagnets},
  author = {Liu, Huiying and Zhao, Jianzhou and Huang, Yue-Xin and Wu, Weikang and Sheng, Xian-Lei and Xiao, Cong and Yang, Shengyuan A.},
  journal = {Phys. Rev. Lett.},
  volume = {127},
  issue = {27},
  pages = {277202},
  numpages = {6},
  year = {2021},
  month = {Dec},
  publisher = {American Physical Society},
  doi = {10.1103/PhysRevLett.127.277202},
  url = {https://link.aps.org/doi/10.1103/PhysRevLett.127.277202}
}

@article{PhysRevLett.129.276601,
  title = {Role of Hidden Spin Polarization in Nonreciprocal Transport of Antiferromagnets},
  author = {Chen, Weizhao and Gu, Mingqiang and Li, Jiayu and Wang, Panshuo and Liu, Qihang},
  journal = {Phys. Rev. Lett.},
  volume = {129},
  issue = {27},
  pages = {276601},
  numpages = {8},
  year = {2022},
  month = {Dec},
  publisher = {American Physical Society},
  doi = {10.1103/PhysRevLett.129.276601},
  url = {https://link.aps.org/doi/10.1103/PhysRevLett.129.276601}
}

@article{PhysRevLett.113.157201,
  title = {Relativistic N\'eel-Order Fields Induced by Electrical Current in Antiferromagnets},
  author = {\ifmmode \check{Z}\else \v{Z}\fi{}elezn\'y, J. and Gao, H. and V\'yborn\'y, K. and Zemen, J. and Ma\ifmmode \check{s}\else \v{s}\fi{}ek, J. and Manchon, Aur\'elien and Wunderlich, J. and Sinova, Jairo and Jungwirth, T.},
  journal = {Phys. Rev. Lett.},
  volume = {113},
  issue = {15},
  pages = {157201},
  numpages = {5},
  year = {2014},
  month = {Oct},
  publisher = {American Physical Society},
  doi = {10.1103/PhysRevLett.113.157201},
  url = {https://link.aps.org/doi/10.1103/PhysRevLett.113.157201}
}

@Article{Wang2023,
author={Wang, Naizhou
and Kaplan, Daniel
and Zhang, Zhaowei
and Holder, Tobias
and Cao, Ning
and Wang, Aifeng
and Zhou, Xiaoyuan
and Zhou, Feifei
and Jiang, Zhengzhi
and Zhang, Chusheng
and Ru, Shihao
and Cai, Hongbing
and Watanabe, Kenji
and Taniguchi, Takashi
and Yan, Binghai
and Gao, Weibo},
title={Quantum-metric-induced nonlinear transport in a topological antiferromagnet},
journal={Nature},
year={2023},
month={Sep},
day={01},
volume={621},
number={7979},
pages={487-492},
issn={1476-4687},
doi={10.1038/s41586-023-06363-3},
url={https://doi.org/10.1038/s41586-023-06363-3}
}

@article{
doi:10.1126/science.adf1506,
author = {Anyuan Gao  and Yu-Fei Liu  and Jian-Xiang Qiu  and Barun Ghosh  and Thaís V. Trevisan  and Yugo Onishi  and Chaowei Hu  and Tiema Qian  and Hung-Ju Tien  and Shao-Wen Chen  and Mengqi Huang  and Damien Bérubé  and Houchen Li  and Christian Tzschaschel  and Thao Dinh  and Zhe Sun  and Sheng-Chin Ho  and Shang-Wei Lien  and Bahadur Singh  and Kenji Watanabe  and Takashi Taniguchi  and David C. Bell  and Hsin Lin  and Tay-Rong Chang  and Chunhui Rita Du  and Arun Bansil  and Liang Fu  and Ni Ni  and Peter P. Orth  and Qiong Ma  and Su-Yang Xu },
title = {Quantum metric nonlinear Hall effect in a topological antiferromagnetic heterostructure},
journal = {Science},
volume = {381},
number = {6654},
pages = {181-186},
year = {2023},
doi = {10.1126/science.adf1506},
URL = {https://www.science.org/doi/abs/10.1126/science.adf1506}}

@article{Buzdin2008,
  author = {Buzdin, A.},
  title = {{Direct Coupling Between Magnetism and Superconducting Current in the Josephson $\varphi_0$ Junction}},
  journal = {Phys. Rev. Lett.},
  volume = {101},
  number = {10},
  pages = {107005},
  year = {2008},
  doi = {10.1103/PhysRevLett.101.107005},
  url = {https://doi.org/10.1103/PhysRevLett.101.107005}
}

@article{Szombati2016,
  author = {Szombati, D. B. and Nadj-Perge, S. and Car, D. and Plissard, S. R. and Bakkers, E. P. A. M. and Kouwenhoven, L. P.},
  title = {{Josephson $\varphi_0$-junction in nanowire quantum dots}},
  journal = {Nat. Phys.},
  volume = {12},
  number = {6},
  pages = {568--572},
  year = {2016},
  doi = {10.1038/nphys3742},
  url = {https://doi.org/10.1038/nphys3742}
}

@article{Strambini2020,
  author = {Strambini, Elia and Iorio, Andrea and Durante, Ofelia and Citro, Roberta and Sanz-Fern{\'a}ndez, Cristina and Guarcello, Claudio and Tokatly, Ilya V. and Braggio, Alessandro and Rocci, Mirko and Ligato, Nadia and Zannier, Valentina and Sorba, Lucia and Bergeret, F. Sebasti{\'a}n and Giazotto, Francesco},
  title = {{A Josephson phase battery}},
  journal = {Nat. Nanotechnol.},
  volume = {15},
  number = {8},
  pages = {656--660},
  year = {2020},
  doi = {10.1038/s41565-020-0712-7},
  url = {https://doi.org/10.1038/s41565-020-0712-7}
}

@article{Daido2022,
  author = {Daido, Akito and Ikeda, Yuhei and Yanase, Youichi},
  title = {{Intrinsic Superconducting Diode Effect}},
  journal = {Phys. Rev. Lett.},
  volume = {128},
  number = {3},
  pages = {037001},
  year = {2022},
  doi = {10.1103/PhysRevLett.128.037001},
  url = {https://doi.org/10.1103/PhysRevLett.128.037001}
}

@article{YuanFu2022,
  author = {Yuan, Noah F. Q. and Fu, Liang},
  title = {{Supercurrent diode effect and finite-momentum superconductors}},
  journal = {Proc. Natl. Acad. Sci. USA},
  volume = {119},
  number = {15},
  pages = {e2119548119},
  year = {2022},
  doi = {10.1073/pnas.2119548119},
  url = {https://doi.org/10.1073/pnas.2119548119}
}

@article{Davydova2022,
  author = {Davydova, Margarita and Prembabu, Saranesh and Fu, Liang},
  title = {{Universal Josephson diode effect}},
  journal = {Sci. Adv.},
  volume = {8},
  number = {23},
  pages = {eabo0309},
  year = {2022},
  doi = {10.1126/sciadv.abo0309},
  url = {https://doi.org/10.1126/sciadv.abo0309}
}

@article{Zhang2022General,
  author = {Zhang, Yi and Gu, Yuhao and Li, Pengfei and Hu, Jiangping and Jiang, Kun},
  title = {{General Theory of Josephson Diodes}},
  journal = {Phys. Rev. X},
  volume = {12},
  number = {4},
  pages = {041013},
  year = {2022},
  doi = {10.1103/PhysRevX.12.041013},
  url = {https://doi.org/10.1103/PhysRevX.12.041013}
}

@article{He2022,
  author = {He, James Jun and Tanaka, Yukio and Nagaosa, Naoto},
  title = {{A phenomenological theory of superconductor diodes}},
  journal = {New J. Phys.},
  volume = {24},
  number = {5},
  pages = {053014},
  year = {2022},
  doi = {10.1088/1367-2630/ac6766},
  url = {https://doi.org/10.1088/1367-2630/ac6766}
}

@article{Hu2023,
  author = {Hu, Jin-Xin and Sun, Zi-Ting and Xie, Ying-Ming and Law, K. T.},
  title = {{Josephson Diode Effect Induced by Valley Polarization in Twisted Bilayer Graphene}},
  journal = {Phys. Rev. Lett.},
  volume = {130},
  number = {26},
  pages = {266003},
  year = {2023},
  doi = {10.1103/PhysRevLett.130.266003},
  url = {https://doi.org/10.1103/PhysRevLett.130.266003}
}

@article{Wang2025,
  author = {Wang, Da and Wang, Qiang-Hua and Wu, Congjun},
  title = {{Current-reversion symmetry breaking and the DC Josephson diode effect}},
  journal = {Sci. Bull.},
  volume = {70},
  number = {24},
  pages = {4181--4186},
  year = {2025},
  doi = {10.1016/j.scib.2025.11.011},
  url = {https://doi.org/10.1016/j.scib.2025.11.011}
}

@article{Baumgartner2022,
  author = {Baumgartner, Christian and Fuchs, Lorenz and Costa, Andreas and Reinhardt, Simon and Gronin, Sergei and Gardner, Geoffrey C. and Lindemann, Tyler and Manfra, Michael J. and Faria Junior, Paulo E. and Kochan, Denis and Fabian, Jaroslav and Paradiso, Nicola and Strunk, Christoph},
  title = {{Supercurrent rectification and magnetochiral effects in symmetric Josephson junctions}},
  journal = {Nat. Nanotechnol.},
  volume = {17},
  number = {1},
  pages = {39--44},
  year = {2022},
  doi = {10.1038/s41565-021-01009-9},
  url = {https://doi.org/10.1038/s41565-021-01009-9}
}

@article{Wu2022,
  author = {Wu, Heng and Wang, Yaojia and Xu, Yuanfeng and Sivakumar, Pranava K. and Pasco, Chris and Filippozzi, Ulderico and Parkin, Stuart S. P. and Zeng, Yu-Jia and McQueen, Tyrel and Ali, Mazhar N.},
  title = {{The field-free Josephson diode in a van der Waals heterostructure}},
  journal = {Nature},
  volume = {604},
  number = {7907},
  pages = {653--656},
  year = {2022},
  doi = {10.1038/s41586-022-04504-8},
  url = {https://doi.org/10.1038/s41586-022-04504-8}
}

@article{Pal2022,
  author = {Pal, Banabir and Chakraborty, Anirban and Sivakumar, Pranava K. and Davydova, Margarita and Gopi, Ajesh K. and Pandeya, Avanindra K. and Krieger, Jonas A. and Zhang, Yang and Date, Mihir and Ju, Sailong and Yuan, Noah and Schr{\"o}ter, Niels B. M. and Fu, Liang and Parkin, Stuart S. P.},
  title = {{Josephson diode effect from Cooper pair momentum in a topological semimetal}},
  journal = {Nat. Phys.},
  volume = {18},
  number = {10},
  pages = {1228--1233},
  year = {2022},
  doi = {10.1038/s41567-022-01699-5},
  url = {https://doi.org/10.1038/s41567-022-01699-5}
}

@article{Bauriedl2022,
  author = {Bauriedl, Lorenz and B{\"a}uml, Christian and Fuchs, Lorenz and Baumgartner, Christian and Paulik, Nicolas and Bauer, Jonas M. and Lin, Kai-Qiang and Lupton, John M. and Taniguchi, Takashi and Watanabe, Kenji and Strunk, Christoph and Paradiso, Nicola},
  title = {{Supercurrent diode effect and magnetochiral anisotropy in few-layer NbSe$_2$}},
  journal = {Nat. Commun.},
  volume = {13},
  number = {1},
  pages = {4266},
  year = {2022},
  doi = {10.1038/s41467-022-31954-5},
  url = {https://doi.org/10.1038/s41467-022-31954-5}
}

@article{Nadeem2023,
  author = {Nadeem, Muhammad and Fuhrer, Michael S. and Wang, Xiaolin},
  title = {{The superconducting diode effect}},
  journal = {Nat. Rev. Phys.},
  volume = {5},
  number = {10},
  pages = {558--577},
  year = {2023},
  doi = {10.1038/s42254-023-00632-w},
  url = {https://doi.org/10.1038/s42254-023-00632-w}
}

@article{Le2024,
  author = {Le, Tian and Pan, Zhiming and Xu, Zhuokai and Liu, Jinjin and Wang, Jialu and Lou, Zhefeng and Yang, Xiaohui and Wang, Zhiwei and Yao, Yugui and Wu, Congjun and Lin, Xiao},
  title = {{Superconducting diode effect and interference patterns in kagome CsV$_3$Sb$_5$}},
  journal = {Nature},
  volume = {630},
  number = {8015},
  pages = {64--69},
  year = {2024},
  doi = {10.1038/s41586-024-07431-y},
  url = {https://doi.org/10.1038/s41586-024-07431-y}
}

@article{Ye2022,
  author = {Ye, Chen and Xie, Xiangnan and Lv, Wenxing and Huang, Ke and Yang, Allen Jian and Jiang, Sicong and Liu, Xue and Zhu, Dapeng and Qiu, Xuepeng and Tong, Mingyu and Zhou, Tong and Hsu, Chuang-Han and Chang, Guoqing and Lin, Hsin and Li, Peisen and Yang, Kesong and Wang, Zhenyu and Jiang, Tian and Renshaw Wang, Xiao},
  title = {{Nonreciprocal Transport in a Bilayer of MnBi$_{2}$Te$_{4}$ and Pt}},
  journal = {Nano Lett.},
  volume = {22},
  number = {3},
  pages = {1366--1373},
  year = {2022},
  doi = {10.1021/acs.nanolett.1c04756},
  url = {https://doi.org/10.1021/acs.nanolett.1c04756}
}

@article{Ren2025,
  author = {Ren, Yafei and Saparov, Daniyar and Niu, Qian},
  title = {{Nonreciprocal Phonons in $\mathcal{PT}$-Symmetric Antiferromagnets}},
  journal = {Phys. Rev. Lett.},
  volume = {134},
  number = {20},
  pages = {206701},
  year = {2025},
  doi = {10.1103/PhysRevLett.134.206701},
  url = {https://doi.org/10.1103/PhysRevLett.134.206701}
}

@misc{Iwata2026,
  author = {Iwata, Takuma and Shiraishi, K. and Aoyama, T. and Senba, D. and Takeda, T. and Fujisawa, Y. and Nurmamat, M. and Nakanishi, K. and Yamagami, K. and Arita, M. and Yamada, T. and Yanagi, Y. and Kimura, A. and Tanida, H. and Kuroda, Kenta},
  title = {{Realization of a parity-violating antiferromagnetic state in LaMnSi}},
  year = {2026},
  eprint = {2605.19891},
  archivePrefix = {arXiv},
  primaryClass = {cond-mat.mtrl-sci},
  doi = {10.48550/arXiv.2605.19891},
  url = {https://arxiv.org/abs/2605.19891}
}

@article{Smejkal2017,
  author = {{\v{S}}mejkal, Libor and {\v{Z}}elezn{\'y}, Jakub and Sinova, Jairo and Jungwirth, Tom{\'a}{\v{s}}},
  title = {{Electric Control of Dirac Quasiparticles by Spin-Orbit Torque in an Antiferromagnet}},
  journal = {Phys. Rev. Lett.},
  volume = {118},
  number = {10},
  pages = {106402},
  year = {2017},
  doi = {10.1103/PhysRevLett.118.106402},
  url = {https://doi.org/10.1103/PhysRevLett.118.106402}
}

@article{Wadley2016,
  author = {Wadley, P. and Howells, B. and {\v{Z}}elezn{\'y}, J. and Andrews, C. and Hills, V. and Campion, R. P. and Nov{\'a}k, V. and Olejn{\'i}k, K. and Maccherozzi, F. and Dhesi, S. S. and Martin, S. Y. and Wagner, T. and Wunderlich, J. and Freimuth, F. and Mokrousov, Y. and Kune{\v{s}}, J. and Chauhan, J. S. and Grzybowski, M. J. and Rushforth, A. W. and Edmonds, K. W. and Gallagher, B. L. and Jungwirth, T.},
  title = {{Electrical switching of an antiferromagnet}},
  journal = {Science},
  volume = {351},
  number = {6273},
  pages = {587--590},
  year = {2016},
  doi = {10.1126/science.aab1031},
  url = {https://doi.org/10.1126/science.aab1031}
}

@article{Godinho2018,
  author = {Godinho, J. and Reichlov{\'a}, H. and Kriegner, D. and Nov{\'a}k, V. and Olejn{\'i}k, K. and Ka{\v{s}}par, Z. and {\v{S}}ob{\'a}{\v{n}}, Z. and Wadley, P. and Campion, R. P. and Otxoa, R. M. and Roy, P. E. and {\v{Z}}elezn{\'y}, J. and Jungwirth, T. and Wunderlich, J.},
  title = {{Electrically induced and detected N{\'e}el vector reversal in a collinear antiferromagnet}},
  journal = {Nat. Commun.},
  volume = {9},
  number = {1},
  pages = {4686},
  year = {2018},
  doi = {10.1038/s41467-018-07092-2},
  url = {https://doi.org/10.1038/s41467-018-07092-2}
}

@article{Baltz2018,
  author = {Baltz, V. and Manchon, A. and Tsoi, M. and Moriyama, T. and Ono, T. and Tserkovnyak, Y.},
  title = {{Antiferromagnetic spintronics}},
  journal = {Rev. Mod. Phys.},
  volume = {90},
  number = {1},
  pages = {015005},
  year = {2018},
  doi = {10.1103/RevModPhys.90.015005},
  url = {https://doi.org/10.1103/RevModPhys.90.015005}
}

@article{Ouassou2023,
  author = {Ouassou, Jabir Ali and Brataas, Arne and Linder, Jacob},
  title = {{dc Josephson Effect in Altermagnets}},
  journal = {Phys. Rev. Lett.},
  volume = {131},
  number = {7},
  pages = {076003},
  year = {2023},
  doi = {10.1103/PhysRevLett.131.076003},
  url = {https://doi.org/10.1103/PhysRevLett.131.076003}
}

@article{BeenakkerVakhtel2023,
  author = {Beenakker, C. W. J. and Vakhtel, T.},
  title = {{Phase-shifted Andreev levels in an altermagnet Josephson junction}},
  journal = {Phys. Rev. B},
  volume = {108},
  number = {7},
  pages = {075425},
  year = {2023},
  doi = {10.1103/PhysRevB.108.075425},
  url = {https://doi.org/10.1103/PhysRevB.108.075425}
}

@article{Lu2024,
  author = {Lu, Bo and Maeda, Kazuki and Ito, Hiroyuki and Yada, Keiji and Tanaka, Yukio},
  title = {{$\varphi$ Josephson Junction Induced by Altermagnetism}},
  journal = {Phys. Rev. Lett.},
  volume = {133},
  number = {22},
  pages = {226002},
  year = {2024},
  doi = {10.1103/PhysRevLett.133.226002},
  url = {https://doi.org/10.1103/PhysRevLett.133.226002}
}

@article{Cheng2024,
  author = {Cheng, Qiang and Mao, Yue and Sun, Qing-Feng},
  title = {{Field-free Josephson diode effect in altermagnet/normal metal/altermagnet junctions}},
  journal = {Phys. Rev. B},
  volume = {110},
  number = {1},
  pages = {014518},
  year = {2024},
  doi = {10.1103/PhysRevB.110.014518},
  url = {https://doi.org/10.1103/PhysRevB.110.014518}
}

@article{ChengHelimagnet2026,
  author = {Cheng, Qiang and Zhuang, Yu-Chen and Sun, Qing-Feng},
  title = {{Helimagnetic Josephson diode effect}},
  journal = {Phys. Rev. B},
  volume = {113},
  number = {17},
  pages = {174518},
  year = {2026},
  doi = {10.1103/82nj-wxgp},
  url = {https://doi.org/10.1103/82nj-wxgp}
}

@misc{LiHuZhang2025,
  author = {Hou, Jin-Xing and Li, Chuang and Hu, Lun-Hui and Zhang, Song-Bo},
  title = {{Field-free Josephson diode and tunable $\varphi_0$-junction in chiral kagome antiferromagnets}},
  year = {2025},
  eprint = {2512.16260},
  archivePrefix = {arXiv},
  primaryClass = {cond-mat.supr-con},
  doi = {10.48550/arXiv.2512.16260},
  url = {https://arxiv.org/abs/2512.16260}
}

@article{Lotfizadeh2024,
  author = {Lotfizadeh, Neda and Schiela, William F. and Pekerten, Baris and Yu, Peng and Elfeky, Bassel Heiba and Strickland, William M. and Matos-Abiague, Alex and Shabani, Javad},
  title = {{Superconducting diode effect sign change in epitaxial Al--InAs Josephson junctions}},
  journal = {Commun. Phys.},
  volume = {7},
  pages = {120},
  year = {2024},
  doi = {10.1038/s42005-024-01618-5},
  url = {https://doi.org/10.1038/s42005-024-01618-5}
}

@article{Mangold2026,
  author = {Mangold, Maximilian and Bauriedl, Lorenz and Berger, Johanna and Yu-Cheng, Chang and Meier, Thomas N. G. and Kronseder, Matthias and Hakonen, Pertti and Back, Christian H. and Strunk, Christoph and Suri, Dhavala},
  title = {{Magnetochiral anisotropy in Josephson diode effect of all-metallic lateral junctions with interfacial Rashba spin-orbit coupling}},
  journal = {Phys. Rev. Lett.},
  volume = {136},
  pages = {206002},
  year = {2026},
  doi = {10.1103/n9g9-99wl},
  url = {https://doi.org/10.1103/n9g9-99wl}
}

@article{Assouline2019,
  author = {Assouline, Alexandre and Feuillet-Palma, Cheryl and Bergeal, Nicolas and Zhang, Tianzhen and Mottaghizadeh, Alireza and Zimmers, Alexandre and Lhuillier, Emmanuel and Eddrie, Mahmoud and Atkinson, Paola and Aprili, Marco and Aubin, Herve},
  title = {{Spin-orbit induced phase-shift in Bi$_2$Se$_3$ Josephson junctions}},
  journal = {Nat. Commun.},
  volume = {10},
  pages = {126},
  year = {2019},
  doi = {10.1038/s41467-018-08022-y},
  url = {https://doi.org/10.1038/s41467-018-08022-y}
}

@article{Lu2023TI,
  author = {Lu, Bo and Ikegaya, Satoshi and Burset, Pablo and Tanaka, Yukio and Nagaosa, Naoto},
  title = {{Tunable Josephson diode effect on the surface of topological insulators}},
  journal = {Phys. Rev. Lett.},
  volume = {131},
  pages = {096001},
  year = {2023},
  doi = {10.1103/PhysRevLett.131.096001},
  url = {https://doi.org/10.1103/PhysRevLett.131.096001}
}

@article{Anh2024,
  author = {Anh, Le Duc and Ishihara, Keita and Hotta, Tomoki and Inagaki, Kohdai and Maki, Hideki and Saeki, Takahiro and Kobayashi, Masaki and Tanaka, Masaaki},
  title = {{Large superconducting diode effect in ion-beam patterned Sn-based superconductor nanowire/topological Dirac semimetal planar heterostructures}},
  journal = {Nat. Commun.},
  volume = {15},
  pages = {8014},
  year = {2024},
  doi = {10.1038/s41467-024-52080-4},
  url = {https://doi.org/10.1038/s41467-024-52080-4}
}

@article{Ma2025NbSe2,
  author = {Ma, Jiaxiang and Wang, Huiyu and Zhuo, Weizhuang and Lei, Bin and Wang, Shuai and Wang, Wenxiang and Chen, Xin-Yu and Wang, Zhen-Yu and Ge, Binghui and Wang, Zhen and Tao, Jing and Jiang, Kun and Xiang, Ziji and Chen, Xian-Hui},
  title = {{Field-free Josephson diode effect in NbSe$_2$ van der Waals junction}},
  journal = {Commun. Phys.},
  volume = {8},
  pages = {125},
  year = {2025},
  doi = {10.1038/s42005-025-02054-9},
  url = {https://doi.org/10.1038/s42005-025-02054-9}
}

@article{Chakraborty2025,
  author = {Chakraborty, Debmalya and Black-Schaffer, Annica M.},
  title = {{Perfect superconducting diode effect in altermagnets}},
  journal = {Phys. Rev. Lett.},
  volume = {135},
  pages = {026001},
  year = {2025},
  doi = {10.1103/cv8s-tk4c},
  url = {https://doi.org/10.1103/cv8s-tk4c}
}

@article{Yang2025NiI2,
  author = {Yang, Hung-Yu and Cuozzo, Joseph J. and Bokka, Anand Johnson and Qiu, Gang and Eckberg, Christopher and Lyu, Yanfeng and Huyan, Shuyuan and Chu, Ching-Wu and Watanabe, Kenji and Taniguchi, Takashi and Wang, Kang L.},
  title = {{Field-resilient supercurrent diode in a multiferroic Josephson junction}},
  journal = {Nat. Commun.},
  volume = {16},
  pages = {9287},
  year = {2025},
  doi = {10.1038/s41467-025-63698-3},
  url = {https://doi.org/10.1038/s41467-025-63698-3}
}

@article{Ghosh2024HighT,
  author = {Ghosh, Sanat and Patil, Vilas and Basu, Amit and Kuldeep and Dutta, Achintya and Jangade, Digambar A. and Kulkarni, Ruta and Thamizhavel, A. and Steiner, Jacob F. and von Oppen, Felix and Deshmukh, Mandar M.},
  title = {{High-temperature Josephson diode}},
  journal = {Nat. Mater.},
  volume = {23},
  pages = {612--618},
  year = {2024},
  doi = {10.1038/s41563-024-01804-4},
  url = {https://doi.org/10.1038/s41563-024-01804-4}
}

@article{Bhowmik2025,
  author = {Bhowmik, Sayak and Samanta, Dibyendu and Nandy, Ashis K. and Saha, Arijit and Ghosh, Sudeep Kumar},
  title = {{Optimizing one dimensional superconducting diodes: interplay of Rashba spin-orbit coupling and magnetic fields}},
  journal = {Commun. Phys.},
  volume = {8},
  pages = {260},
  year = {2025},
  doi = {10.1038/s42005-025-02044-x},
  url = {https://doi.org/10.1038/s42005-025-02044-x}
}

@article{Zhu2025Andreev,
  author = {Zhu, Shang and Ma, Yiwen and He, Jiangbo and Yang, Xiaozhou and Jia, Zhongmou and Wei, Min and Jiao, Yiping and He, Jiezhong and Zhuo, Enna and Cao, Xuewei and Tong, Bingbing and Dou, Ziwei and Li, Peiling and Shen, Jie and Song, Xiaohui and Lyu, Zhaozheng and Liu, Guangtong and Pan, Dong and Zhao, Jianhua and Lu, Bo and Lu, Li and Qu, Fanming},
  title = {{Josephson diode effect in nanowire-based Andreev molecules}},
  journal = {Commun. Phys.},
  volume = {8},
  pages = {330},
  year = {2025},
  doi = {10.1038/s42005-025-02237-4},
  url = {https://doi.org/10.1038/s42005-025-02237-4}
}

@article{Tanaka2022DwaveTI,
  author = {Tanaka, Yukio and Lu, Bo and Nagaosa, Naoto},
  title = {{Theory of giant diode effect in $d$-wave superconductor junctions on the surface of a topological insulator}},
  journal = {Phys. Rev. B},
  volume = {106},
  pages = {214524},
  year = {2022},
  doi = {10.1103/PhysRevB.106.214524},
  url = {https://doi.org/10.1103/PhysRevB.106.214524}
}

@article{Zeng2025TiltedDirac,
  author = {Zeng, W.},
  title = {{Transverse Josephson diode effect in tilted Dirac systems}},
  journal = {Phys. Rev. Lett.},
  volume = {134},
  pages = {176002},
  year = {2025},
  doi = {10.1103/PhysRevLett.134.176002},
  url = {https://doi.org/10.1103/PhysRevLett.134.176002}
}

@article{McFarlane2026,
  author = {McFarlane, Emily C. and Sanna, Antonio and Gilbert, Matthew J. and Krieger, Jonas A. and Date, Mihir and Domaine, Gabriele and Pal, Banabir and Chakraborty, Anirban and Sivakumar, Pranava K. and Constantinou, Procopios C. and Hartl, Anna and Della Valle, Enrico G. and Pellegrini, Camilla and Strocov, Vladimir N. and Parkin, Stuart S. P. and Schr{\"o}ter, Niels B. M.},
  title = {{Van Hove singularities, superconductivity, and the Josephson diode effect in NiTe$_2$ and PdTe$_2$}},
  journal = {Phys. Rev. Lett.},
  volume = {136},
  pages = {086401},
  year = {2026},
  doi = {10.1103/hp1t-zd7y},
  url = {https://doi.org/10.1103/hp1t-zd7y}
}

@article{DiezMerida2023,
  author = {D{\'i}ez-M{\'e}rida, J. and D{\'i}ez-Carl{\'o}n, A. and Yang, S. Y. and Xie, Y.-M. and Gao, X.-J. and Senior, J. and Watanabe, K. and Taniguchi, T. and Lu, X. and Higginbotham, A. P. and Law, K. T. and Efetov, Dmitri K.},
  title = {{Symmetry-broken Josephson junctions and superconducting diodes in magic-angle twisted bilayer graphene}},
  journal = {Nat. Commun.},
  volume = {14},
  pages = {2396},
  year = {2023},
  doi = {10.1038/s41467-023-38005-7},
  url = {https://doi.org/10.1038/s41467-023-38005-7}
}

@article{Volkov2024Twisted,
  author = {Volkov, Pavel A. and Lantagne-Hurtubise, {\'E}tienne and Tummuru, Tarun and Plugge, Stephan and Pixley, J. H. and Franz, Marcel},
  title = {{Josephson diode effects in twisted nodal superconductors}},
  journal = {Phys. Rev. B},
  volume = {109},
  pages = {094518},
  year = {2024},
  doi = {10.1103/PhysRevB.109.094518},
  url = {https://doi.org/10.1103/PhysRevB.109.094518}
}

@article{Wang2026TwistedFeSe,
  author = {Wang, Juyuan and Wei, Wei and Pan, Chuandi and Wang, Hengning and Wang, Chunsheng and Sun, Yue and Shi, Zhixiang and Niu, Qun and Zheng, Guolin and Tian, Mingliang},
  title = {{Field-free superconducting diode effect in 45$^{\circ}$-twisted FeSe van der Waals Josephson junctions}},
  journal = {Materials},
  volume = {19},
  pages = {972},
  year = {2026},
  doi = {10.3390/ma19050972},
  url = {https://doi.org/10.3390/ma19050972}
}

@article{Lou2026Kagome,
  author = {Lou, Han-Xin and Chen, Jing-Jing and Ye, Xing-Guo and Tan, Zhen-Bing and Wang, An-Qi and Yin, Qing and Liao, Xin and Fang, Jing-Zhi and Liu, Xing-Yu and He, Yi-Lin and Zhang, Zhen-Tao and Li, Chuan and Wei, Zhong-Ming and Ma, Xiu-Mei and Yu, Da-Peng and Liao, Zhi-Min},
  title = {{Quantized radio-frequency rectification in a kagome superconductor Josephson diode}},
  journal = {Nat. Nanotechnol.},
  volume = {21},
  pages = {554--560},
  year = {2026},
  doi = {10.1038/s41565-025-02120-x},
  url = {https://doi.org/10.1038/s41565-025-02120-x}
}

@article{Su2024,
  author = {Su, Haitian and Wang, Ji-Yin and Gao, Han and Luo, Yi and Yan, Shili and Wu, Xingjun and Li, Guoan and Shen, Jie and Lu, Li and Pan, Dong and Zhao, Jianhua and Zhang, Po and Xu, H. Q.},
  title = {{Microwave-assisted unidirectional superconductivity in Al--InAs nanowire--Al junctions under magnetic fields}},
  journal = {Phys. Rev. Lett.},
  volume = {133},
  pages = {087001},
  year = {2024},
  doi = {10.1103/PhysRevLett.133.087001},
  url = {https://doi.org/10.1103/PhysRevLett.133.087001}
}

@article{Legg2023Parity,
  author = {Legg, Henry F. and Laubscher, Katharina and Loss, Daniel and Klinovaja, Jelena},
  title = {{Parity-protected superconducting diode effect in topological Josephson junctions}},
  journal = {Phys. Rev. B},
  volume = {108},
  pages = {214520},
  year = {2023},
  doi = {10.1103/PhysRevB.108.214520},
  url = {https://doi.org/10.1103/PhysRevB.108.214520}
}

@article{CostaFabian2023,
  author = {Costa, Andreas and Fabian, Jaroslav and Kochan, Denis},
  title = {{Microscopic study of the Josephson supercurrent diode effect in Josephson junctions based on two-dimensional electron gas}},
  journal = {Phys. Rev. B},
  volume = {108},
  pages = {054522},
  year = {2023},
  doi = {10.1103/PhysRevB.108.054522},
  url = {https://doi.org/10.1103/PhysRevB.108.054522}
}

@misc{SM,
  title = {{Supplemental Material for ``N{\'e}el-Vector Control of the Josephson Diode Effect in $\mathcal{PT}$-symmetric Antiferromagnets''}},
  note = {See Supplemental Material for the numerical calculations,electric-field control,channel-resolved Matsubara theory, the exact Green-function reduction, and the magnetic point group classification.The parameters used in all figures are described in the SM.}
}

@article{Xiao2023MagneticSHG,
  author = {Xiao, Rui-Chun and Shao, Ding-Fu and Gan, Wei and Wang, Huan-Wen
            and Han, Hui and Sheng, Z. G. and Zhang, Changjin and Jiang, Hua
            and Li, Hui},
  title = {Classification of second harmonic generation effect in magnetically ordered materials},
  journal = {npj Quantum Mater.},
  volume = {8},
  pages = {62},
  year = {2023},
  doi = {10.1038/s41535-023-00594-3}
}

@article{Hu2026PTJosephson,
  author  = {Hu, Jin-Xin and Hu, Mengli and Xie, Ying-Ming and Law, K. T.},
  title   = {Electrically controlled {0--$\pi$} oscillations and antiferromagnetic {Josephson} spin valve with {$PT$} symmetry},
  journal = {npj Quantum Materials},
  year    = {2026},
  month   = jul,
  doi     = {10.1038/s41535-026-00920-5},
  url     = {https://doi.org/10.1038/s41535-026-00920-5}
}

\clearpage
\onecolumngrid
\begin{bibunit}[apsrev4-2]

\makeatletter
\let\@FMN@list\@empty
\renewcommand{\@biblabel}[1]{[S#1]}
\makeatother

\renewcommand{\citenumfont}[1]{S#1}
\renewcommand{\bibnumfmt}[1]{[S#1]}

\SupplementTOCtrue
\begin{center}
\begin{large}
\textbf{Supplemental Material for ``Néel-Vector Control of the Josephson Diode Effect in
$\mathcal{PT}$-symmetric Antiferromagnets''}
\end{large}
\end{center}

\setcounter{section}{0}
\setcounter{subsection}{0}
\setcounter{figure}{0}
\setcounter{table}{0}
\setcounter{equation}{0}
\setcounter{secnumdepth}{2}
\setcounter{tocdepth}{2}
\renewcommand{\thesection}{\Alph{section}}
\renewcommand{\thesubsection}{\thesection\arabic{subsection}}
\renewcommand{\thefigure}{S\arabic{figure}}
\renewcommand{\thetable}{S\arabic{table}}
\renewcommand{\theequation}{S\arabic{equation}}
\renewcommand{\theHsection}{supp.\arabic{section}}
\renewcommand{\theHsubsection}{\theHsection.\arabic{subsection}}
\renewcommand{\theHfigure}{supp.\arabic{figure}}
\renewcommand{\theHtable}{supp.\arabic{table}}
\renewcommand{\theHequation}{supp.\arabic{equation}}

\tableofcontents

\section{Numerical calculations}

This section describes the lattice calculation of the Josephson current and verifies the symmetry-forbidden diode response for a N\'eel vector along $y$.

\subsection{BdG diagonalization and current extraction}

We implement the lattice BdG Hamiltonian of the main text in the balanced AB--AB junction geometry, with equal numbers of $A$ and $B$ sites in the normal region. The junction is periodic along $x$ and finite along the transport direction $y$, so $k_x$ labels independent transverse channels. The staggered exchange is confined to the normal region, and the superconducting regions have prescribed pair potentials with phases $\pm\varphi/2$; the pairing amplitude is not determined self-consistently.

For direct BdG validation at a sampled pair $(k_x,\varphi)$, we diagonalize the complete finite-$y$ Hamiltonian $\mathcal H_{\rm BdG}(k_x,\varphi)$ and obtain its eigenvalues $E_n(k_x,\varphi)$. Up to phase-independent terms, the free energy averaged over $N_k$ transverse momenta and the corresponding current are
\begin{equation}
\begin{aligned}
F_T(\varphi)
&=-\frac{k_BT}{2N_k}\sum_{k_x,n}
\ln\!\left[2\cosh\frac{E_n(k_x,\varphi)}{2k_BT}\right],\\
F_0(\varphi)
&=-\frac{1}{4N_k}\sum_{k_x,n}|E_n(k_x,\varphi)|,\\
I(\varphi)&=\frac{2e}{\hbar}\frac{\partial F_T(\varphi)}{\partial\varphi}.
\end{aligned}
\label{eq:s-numerical-current}
\end{equation}
The sum includes the complete positive- and negative-energy spectrum, accounting for the BdG redundancy through the prefactors above. All physical multiplicities are already included in this spectrum, so no additional degeneracy factor is applied. For a single-channel calculation, the transverse-momentum average is replaced by the spectrum at the specified $k_x$. Currents are expressed in units of $e\Delta_0/\hbar$.

We sample $\varphi$ uniformly over $[0,2\pi)$ with $N_\varphi$ points. Figure~3(a) uses direct BdG diagonalization; the updated Figs.~1(c), 2(a--c), and 3(c), as well as Fig.~\ref{fig:s-neel-y-cpr}, use the exact finite-electrode Green determinant of the same BdG matrix to evaluate the current on the phase mesh. The directional critical currents and diode efficiency are defined by
\begin{equation}
I_c^+=\max_\varphi I(\varphi),\qquad
I_c^-=\min_\varphi I(\varphi),\qquad
\eta=\frac{I_c^+-|I_c^-|}{I_c^++|I_c^-|}.
\label{eq:s-numerical-eta}
\end{equation}
Critical-current extrema are refined by continuous optimization of a periodic current spline rather than being restricted to the sampled phase points.

Unless stated otherwise, the reference parameters are $t=1$, $t'=0.1t$, $\Delta_0=0.075t$, $\mu=-1.1t$, $J_n=0.4t$, $\lambda_N=0.25t$, $d_N=12$, $W_S=20$, $J_S=\lambda_S=0$, and zero SOC on NS interface bonds. Here $d_N$ and $W_S$ specify the numbers of unit-cell layers in the normal region and each superconducting region, respectively.

Figure~1(c) uses the reference parameters with $N_k=2561$, $N_\varphi=1441$, and $N_\omega=768$ at $T=0$.

Figure~2(a) uses $N_k=321$, $N_\varphi=1441$, and $N_\omega=256$ at $T=0$. The normalized product takes the values $\lambda_N J_n/(\lambda_0J_0)=0$, $0.4$, $0.8$, $1$, $1.2$, $1.6$, and $2$. The nonzero-product cases have $J_n/\lambda_N=1.6$, while both parameters vanish at zero product.

For Fig.~2(b), we use $N_k=1281$, $N_\varphi=1441$, and $N_\omega=384$ at $T=0$. Defining $x=\lambda_N J_n/(\lambda_0J_0)$ with $J_0=0.4t$ and $\lambda_0=0.25t$.

For Fig.~2(c), we fix $J_n=0.4t$ and $\lambda_N=0.25t$ and rotate $\boldsymbol n=(\cos\theta,\sin\theta,0)$ at constant exchange magnitude. The calculations use $N_k=1281$, $N_\varphi=1441$, and $N_\omega=384$ at $T=0$. 

Figure~3(a,b) uses the reference parameters at fixed $k_x=0.3\pi$ and $k_BT=0.05\Delta_0$. The BdG CPR is calculated with $N_\varphi=1441$, and the Matsubara calculation retains 768 positive frequencies. The harmonic analysis in Fig.~3(b) uses these same CPRs. For the channel-resolved efficiency in Fig.~3(c), $N_k=2561$ transverse momenta are sampled, with $N_\varphi=11521$ and $N_\omega=768$ for each channel at $T=0$. The Green-function method used in Fig.~3(d) is described separately in the section on the exact Green-function reduction.

\subsection{CPR for a N\'eel vector along $y$}

A N\'eel vector along $y$ restores the current-reversing magnetic mirror $\mathcal M_y$, which requires $F_T(\varphi)=F_T(-\varphi)$ and hence $I(\varphi)=-I(-\varphi)$. The positive and negative critical-current magnitudes must therefore be equal.

We test this constraint using the same C parameters as in Fig.~1(c), namely $\mu=-1.1t$, $J_n=0.4t$, and $\lambda_N=0.25t$, but with $\boldsymbol n\parallel\hat y$. The calculation is performed at $T=0$, with $N_k=1281$, $N_\varphi=1441$, and $N_\omega=384$.

\begin{figure}[b]
\centering
\includegraphics[width=0.54\textwidth]{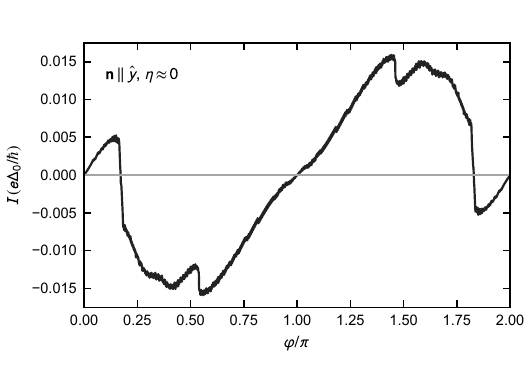}
\caption{\label{fig:s-neel-y-cpr}
Transverse-momentum-averaged finite-SNS CPR for $\boldsymbol n\parallel\hat y$ at $\mu=-1.1t$, $J_n=0.4t$, $\lambda_N=0.25t$, $t'=0.1t$, $\Delta_0=0.075t$, $d_N=12$, and $W_S=20$, with $J_S=\lambda_S=0$ and zero interface SOC.
The calculation uses $T=0$, $N_k=1281$ uniformly sampled transverse momenta, $N_\varphi=1441$ phase points, and $N_\omega=384$ Matsubara integration nodes. The CPR is obtained without imposing its odd symmetry.}
\end{figure}

The resulting CPR is displayed in Fig.~\ref{fig:s-neel-y-cpr}. No odd-symmetry constraint is imposed on the calculated current. We obtain $|\eta|<1.0\times10^{-11}$, and the normalized residual $\max_\varphi|I(\varphi)+I(-\varphi)|/\max_\varphi|I(\varphi)|$ is below $6.0\times10^{-9}$. These residuals quantify the numerical preservation of the mirror constraint and confirm the absence of a diode response for this orientation.

\section{Optional electric-field control}

The Josephson diode effect discussed in the main text does not require an
external electric field.  Nevertheless, an electric-field-induced potential
difference between the two inversion-partner sublattices provides an
additional control parameter.  We include it through
\begin{equation}
 \mathcal H_v=-v\rho_z\tau_z,
\label{eq:s-electric-field}
\end{equation}
so that the normal-state on-site energies of the $A$ and $B$ sublattices are
shifted by $-v$ and $+v$, respectively.  The term is included throughout the
gated material, whereas the staggered exchange remains confined to the normal
weak link, as in the main-text model.

The sublattice potential is odd under $\PT$:
$\PT\,\mathcal H_v\,(\PT)^{-1}=-\mathcal H_v$,
because $\PT$ exchanges the two sublattices and leaves $\tau_z$ unchanged.
Thus, a fixed nonzero $v$ explicitly breaks the $\PT$ symmetry of the
ungated normal-state model, so its twofold band degeneracy is no longer
protected by this symmetry. Electric-field control therefore extends the model
away from the $\PT$-symmetric limit; it is not required for the diode
response already present at $v=0$.

\begin{figure}[t]
\centering
\includegraphics[width=0.94\textwidth]
{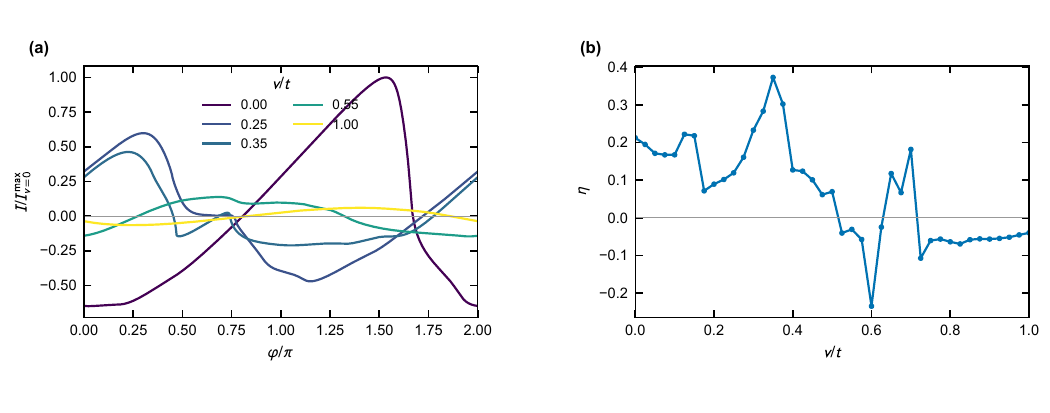}
\caption{\label{fig:s-electric-field}
Electric-field control in the balanced AB--AB SNS junction at
$\mu=-1.1t$, $J_n=0.4t$, and $\lambda_N=0.25t$, with $t'=0.1t$, $\Delta_0=0.075t$, $d_N=12$, $W_S=20$, $J_S=\lambda_S=0$, and zero interface SOC.
(a) CPRs for representative staggered potentials $v$.  All curves use the
same current normalization $I^{\max}_{v=0}$, so the suppression of the total current at large
$v$ remains visible.
(b) Diode efficiency as a function of $v$.  The calculation uses
$N_k=321$, $N_\varphi=1441$, $N_\omega=384$, and 41 uniformly spaced values of $v/t$ between
0 and 1; critical currents are extracted from periodic cubic interpolation
of the calculated current.}
\end{figure}

Figure~\ref{fig:s-electric-field} shows that $v$ does not merely produce a
rigid displacement of the CPR.  It changes both the phase structure and the
overall current scale [Fig.~\ref{fig:s-electric-field}(a)], demonstrating that
the sublattice potential reweights the transverse transport channels and their
harmonics.  Consequently, the diode efficiency is strongly nonmonotonic
[Fig.~\ref{fig:s-electric-field}(b)].  For the reference parameters it reaches
$\eta\simeq0.373$ near $v=0.35t$ and reverses sign as $v$ is increased further.
The finite value at $v=0$ confirms that the electric field is not the origin
of the diode response.  Rather, it is an optional knob that modifies the
coherent channel sum and can tune both the magnitude and polarity of the
effect.

\section{Channel-resolved scattering theory on the Matsubara axis}

This section derives the channel CPR from the two oppositely directed
Andreev loops and identifies the microscopic condition for a finite
single-channel diode response.

\begin{figure}[t]
\centering
\includegraphics[width=0.82\textwidth]{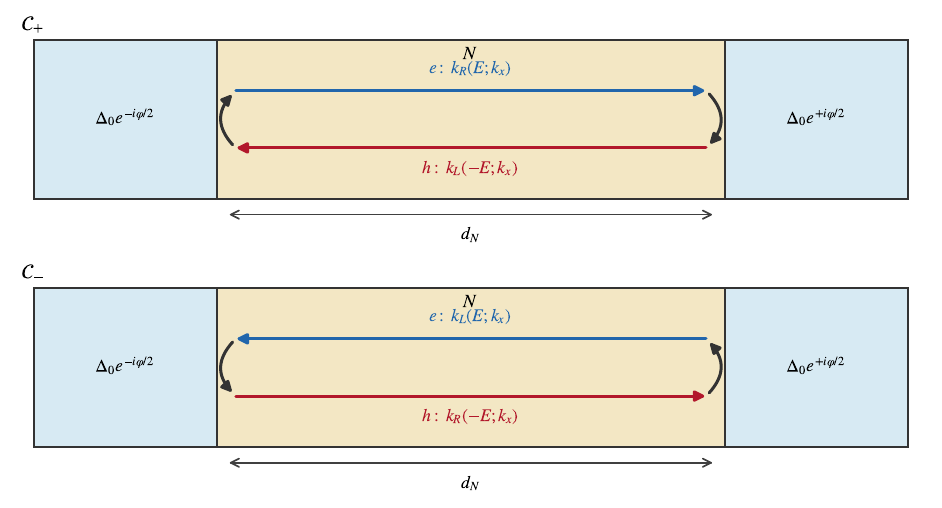}
\caption{\label{fig:s-andreev-loops}
The two oppositely directed Andreev loops at fixed transverse momentum $k_x$.
The loop $\mathcal C_+$ consists of a right-moving electron and a left-moving
hole and acquires the superconducting phase $+\varphi$; $\mathcal C_-$
contains the opposite propagation sequence and acquires $-\varphi$.
The normal-region propagation phases are $\Theta_+(E)$ and $\Theta_-(E)$ in
Eq.~\eqref{eq:s-loop-phase}.}
\end{figure}

At fixed $k_x$, we consider the active bonding branch and suppress the
$k_x$ label until it is needed again.  Its right- and left-moving roots are
defined by
\begin{align}
\varepsilon_-[k_x,k_R(E)]&=E,
&v_y[k_R(E)]&>0,\nonumber\\
\varepsilon_-[k_x,k_L(E)]&=E,
&v_y[k_L(E)]&<0.
\label{eq:s-roots}
\end{align}
For $d_N$ normal layers, the propagation distance between the reference
planes of the first and last normal layers is $L_N=(d_N-1)a$, where $a$ is
the layer spacing.  With momenta measured in units of $a^{-1}$, the two
oppositely directed Andreev loops have the propagation phases
\begin{align}
\Theta_+(E)
&=L_N\left[k_R(E)+k_L(-E)\right],\nonumber\\
\Theta_-(E)
&=-L_N\left[k_R(-E)+k_L(E)\right].
\label{eq:s-loop-phase}
\end{align}
The two closed trajectories are illustrated in
Fig.~\ref{fig:s-andreev-loops}.  The blue and red paths denote electron and
hole propagation, respectively, while the short black arrows denote Andreev
conversion at a transparent NS interface.  Ordinary normal reflection is not
included at this stage.

The arguments $\pm E$ arise because a hole at BdG energy $E$ is the absence
of an electron at energy $-E$.  Direct substitution gives
\begin{equation}
\Theta_-(E)=-\Theta_+(-E).
\label{eq:s-loop-relation}
\end{equation}
The validity of this relation requires spin degeneracy in the system. For
transparent NS interfaces, the two loop quantization conditions can be
written as
\begin{align}
-2\arccos\frac{E}{\Delta_0}
+\varphi+\Theta_+(E)&=2\pi l,\nonumber\\
-2\arccos\frac{E}{\Delta_0}
-\varphi+\Theta_-(E)&=2\pi l,
\qquad l\in\mathbb Z .
\label{eq:s-quantization}
\end{align}
The first term is the phase accumulated in two Andreev reflections,
$\pm\varphi$ is the superconducting phase acquired by the two opposite
loops, and $\Theta_\pm$ is their normal-region propagation phase.

We now continue the same roots from real energy to the positive Matsubara
axis, $E=i\omega_n$.  Writing
\begin{equation}
\Theta_+(i\omega_n)
=\alpha(\omega_n)+i\beta(\omega_n),
\label{eq:s-alpha-beta}
\end{equation}
and using Eq.~(\ref{eq:s-loop-relation}) together with
$\Theta_+(-i\omega_n)=\Theta_+^*(i\omega_n)$ gives
\begin{equation}
\Theta_-(i\omega_n)
=-\alpha(\omega_n)+i\beta(\omega_n).
\label{eq:s-opposite-loop-matsubara}
\end{equation}
The real part $\alpha$ is therefore the propagation phase of the two loops,
with opposite signs for opposite orientations, whereas the common imaginary
part $\beta$ attenuates both loops.  The attenuation factor of a complete
Andreev round trip is
\begin{equation}
\rho(\omega_n)
=\exp\left[
-2\operatorname{arsinh}\frac{\omega_n}{\Delta_0}
-\beta(\omega_n)
\right].
\label{eq:s-rho-meaning}
\end{equation}
The first term in the exponent is the decay supplied by the two Andreev
reflections and the second is the decay accumulated during propagation
through the normal region.

Equations~(\ref{eq:s-quantization})--(\ref{eq:s-rho-meaning}) give the
complete two-loop determinant on the Matsubara axis,
\begin{align}
\mathcal D(i\omega_n,\varphi)
&=\left[1+\rho e^{i(\varphi+\alpha)}\right]
  \left[1+\rho e^{-i(\varphi+\alpha)}\right]\nonumber\\
&=1+2\rho\cos(\varphi+\alpha)+\rho^2,
\label{eq:s-channel-determinant}
\end{align}
where the $\omega_n$ arguments of $\rho$ and $\alpha$ are suppressed only
inside the same equation.  After positive and negative Matsubara frequencies
are combined, the phase-dependent thermodynamic potential is
\begin{equation}
\Omega_{k_x}(\varphi)
=-\frac{g_{k_x}}{\beta_T}
\sum_{\omega_n>0}\ln\mathcal D(i\omega_n,\varphi),
\qquad g_{k_x}=2,
\label{eq:s-channel-free-energy}
\end{equation}
up to a $\varphi$-independent constant.  Here $\beta_T=(k_BT)^{-1}$ and
$g_{k_x}=2$ accounts for the twofold $\mathcal{PT}$ degeneracy at $v=0$.  Using
$I_{k_x}=(2e/\hbar)\partial_\varphi\Omega_{k_x}$ yields
\begin{equation}
I_{k_x}^{\mathrm M}(\varphi)
=\frac{4eg_{k_x}}{\hbar\beta_T}
\sum_{\omega_n>0}
\frac{\rho(\omega_n)\sin[\varphi+\alpha(\omega_n)]}
{1+2\rho(\omega_n)\cos[\varphi+\alpha(\omega_n)]
+\rho^2(\omega_n)}.
\label{eq:s-matsubara-current}
\end{equation}

This expression separates the two ingredients of the channel response.
The real phase $\alpha(\omega_n)$ fixes the phase center of each Matsubara
contribution, while $\rho(\omega_n)$ fixes its weight.  If
$\alpha(\omega_n)=\alpha_0$ is independent of frequency, every term is odd
about the same translated origin:
\begin{equation}
I_{k_x}^{\mathrm M}(-\alpha_0+\chi)
=-I_{k_x}^{\mathrm M}(-\alpha_0-\chi).
\label{eq:s-common-center}
\end{equation}
The channel can then be a $\varphi_0$ junction, but its positive and negative
critical-current magnitudes are equal, so it has no diode effect.  The
frequency dependence of $\rho$ alone cannot change this conclusion because
it changes only the weights of functions having the same phase center.

The first term that makes the real phase frequency dependent is exposed by
expanding the propagation phase near zero energy:
\begin{equation}
\Theta_+(E)=\Theta_0+\tau E+qE^2+\mathcal O(E^3),
\label{eq:s-theta-expansion}
\end{equation}
where
\begin{align}
\Theta_0
&=L_N\left[k_R(0)+k_L(0)\right],\nonumber\\
\tau
&=L_N\left(\frac{1}{v_R}-\frac{1}{v_L}\right),\nonumber\\
q
&=-\frac{L_N}{2}
\left[
\frac{\varepsilon_-''(k_R)}{v_R^3}
+\frac{\varepsilon_-''(k_L)}{v_L^3}
\right]_{E=0}.
\label{eq:s-expansion-coefficients}
\end{align}
Here $v_{R,L}=\partial_{k_y}\varepsilon_-(k_x,k_y)|_{k_{R,L}(0)}$.
The constant $\Theta_0$ supplies the zero-energy anomalous phase.  The linear
term $\tau E$ measures the dynamical propagation time; after
$E=i\omega_n$ it becomes purely imaginary and therefore modifies the
attenuation.  By contrast, the curvature term becomes real:
\begin{equation}
\Theta_+(i\omega_n)
=\Theta_0+i\tau\omega_n-q\omega_n^2+\cdots,
\qquad
\alpha(\omega_n)
=\Theta_0-q\omega_n^2+\cdots.
\label{eq:s-alpha-expansion}
\end{equation}
Thus the quadratic energy dependence shifts the phase centers of different
Matsubara contributions by different amounts.  This effect is absent if the
propagation phase is truncated at linear order.

The resulting phase mismatch is seen directly by expanding
Eq.~(\ref{eq:s-matsubara-current}) into harmonics:
\begin{equation}
\frac{\rho\sin x}{1+2\rho\cos x+\rho^2}
=\sum_{m=1}^{\infty}(-1)^{m+1}\rho^m\sin(mx).
\label{eq:s-fourier-identity}
\end{equation}
Accordingly,
\begin{align}
I_{k_x}^{\mathrm M}(\varphi)
&=\operatorname{Im}\sum_{m\geq1}
e^{im\varphi}\mathcal C_{m,k_x},\nonumber\\
\mathcal C_{m,k_x}
&=\frac{4eg_{k_x}}{\hbar\beta_T}(-1)^{m+1}
\sum_{\omega_n>0}\rho^m(\omega_n)e^{im\alpha(\omega_n)}.
\label{eq:s-channel-fourier-coefficient}
\end{align}
Using Eq.~(\ref{eq:s-alpha-expansion}), define
\begin{equation}
R_m=\sum_{\omega_n>0}\rho^m(\omega_n),
\qquad
\langle\omega^2\rangle_m
=\frac{\sum_{\omega_n>0}\omega_n^2\rho^m(\omega_n)}
{\sum_{\omega_n>0}\rho^m(\omega_n)}.
\label{eq:s-frequency-moment}
\end{equation}
To first order in $q$,
\begin{equation}
\mathcal C_{m,k_x}
\simeq
\frac{4eg_{k_x}}{\hbar\beta_T}(-1)^{m+1}
e^{im\Theta_0}R_m
\left[1-imq\langle\omega^2\rangle_m\right].
\label{eq:s-coefficient-expansion}
\end{equation}
Its phase is therefore
\begin{equation}
\theta_{m,k_x}
\simeq m\Theta_0-mq\langle\omega^2\rangle_m
\pmod{\pi}.
\label{eq:s-harmonic-phase}
\end{equation}
Taking the phase of the first harmonic as the reference, the mismatch of the
$m$th harmonic is
\begin{equation}
\delta_{m,k_x}
\equiv\theta_{m,k_x}-m\theta_{1,k_x}
\simeq
-mq\left(
\langle\omega^2\rangle_m-\langle\omega^2\rangle_1
\right)
\pmod{\pi}.
\label{eq:s-general-harmonic-unlocking}
\end{equation}
Equation~(\ref{eq:s-general-harmonic-unlocking}) fixes the harmonic-order
dependence once the low-frequency form of the Matsubara weight is specified.
From Eqs.~(\ref{eq:s-rho-meaning}) and
(\ref{eq:s-alpha-expansion}),
\begin{equation}
\beta(\omega)=\tau\omega+\mathcal O(\omega^3),
\qquad
-\ln\rho(\omega)=a\omega+\mathcal O(\omega^3),
\qquad
a=\frac{2}{\Delta_0}+\tau .
\label{eq:s-low-frequency-rho}
\end{equation}
We assume the generic attenuating case $a>0$.  At sufficiently low
temperature, the Matsubara sum may be replaced by an integral, provided the
characteristic frequencies selected by $\rho^m$ remain inside the low-energy
window of Eq.~(\ref{eq:s-low-frequency-rho}).  Since
$\rho^m(\omega)\simeq\exp(-ma\omega)$, the weighted moment becomes
\begin{align}
\langle\omega^2\rangle_m
&\simeq
\frac{\displaystyle\int_0^\infty d\omega\,
\omega^2e^{-ma\omega}}
{\displaystyle\int_0^\infty d\omega\,e^{-ma\omega}}
=\frac{2}{a^2m^2}.
\label{eq:s-low-temperature-moment}
\end{align}
Substitution into Eq.~(\ref{eq:s-general-harmonic-unlocking}) gives the
low-temperature asymptotic relation
\begin{equation}
\delta_{m,k_x}
\simeq
\Delta_{\theta,k_x}\left(m-\frac{1}{m}\right),
\qquad
\Delta_{\theta,k_x}=\frac{2q}{a^2}
\label{eq:s-m-minus-inverse-m}
\end{equation}
to first order in the curvature coefficient $q$.  In particular,
$\delta_{2,k_x}\simeq3q/a^2$.  The neighboring phase spacings obey
$\delta_{m+1,k_x}-\delta_{m,k_x}
=\Delta_{\theta,k_x}[1+1/(m(m+1))]$ and rapidly approach a constant.
Therefore, over a finite range of harmonic orders, the
$m-1/m$ dependence can appear nearly linear even though it is distinct from
an exact $m-1$ law.  Finite temperature, higher-order terms in
$-\ln\rho(\omega)$, and higher powers of $q$ generate systematic deviations
from Eq.~(\ref{eq:s-m-minus-inverse-m}).

\begin{figure}[t]
\centering
\includegraphics[width=0.98\textwidth]{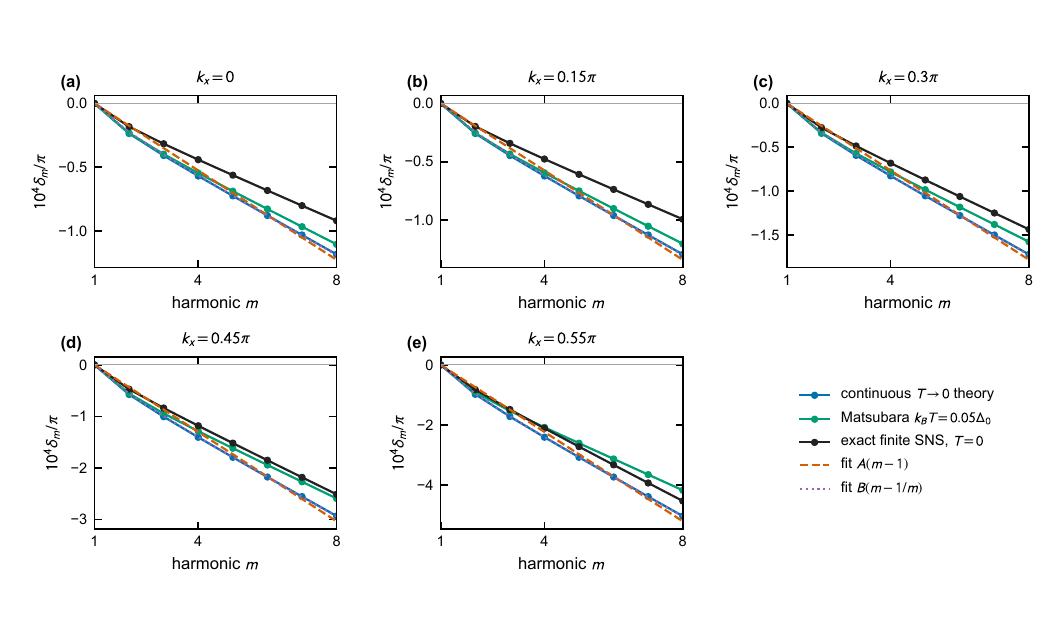}
\caption{\label{fig:s-harmonic-unlocking}
Harmonic phase mismatch $\delta_{m,k_x}/\pi$ for
$k_x/\pi=0$, $0.15$, $0.30$, $0.45$, and $0.55$.
Blue, green, and black curves show the continuous $T\to0$ scattering theory,
the discrete Matsubara calculation at $k_BT=0.05\Delta_0$, and the full
zero-temperature BdG calculation, respectively.  The dashed and dotted
curves are the $A(m-1)$ and $B(m-1/m)$ fits to the continuous theory; the
corresponding fits to the other two data sets are assessed through their
residuals in the text and are omitted to avoid clutter.  All panels use
$\mu=-1.1t$, $J_n=0.4t$, $\lambda_N=0.25t$, $t'=0.1t$, $\Delta_0=0.075t$, $d_N=12$, and $W_S=20$, with zero SOC in the superconductors and on NS interface bonds.}
\end{figure}

Figure~\ref{fig:s-harmonic-unlocking} tests this harmonic-order dependence
for five representative transverse channels.  For each data set, the
coefficients multiplying $m-1$ and $m-1/m$ are fitted independently over
$m=2,\ldots,8$.  The residual ratio
$\mathrm{RSS}_{m-1}/\mathrm{RSS}_{m-1/m}$ lies between $71.1$ and $323$ for the
continuous low-temperature theory, between $8.75$ and $12.86$ for the discrete
Matsubara calculation at $k_BT=0.05\Delta_0$, and between $15.67$ and $640.36$ for
the full BdG result.  Thus all three calculations favor the predicted
$m-1/m$ dependence over a strictly linear $m-1$ law, although their fitted
prefactors need not coincide.

More generally, different harmonics sample the Matsubara spectrum with the
different weights $\rho^m$, so their frequency moments are unequal.
Consequently, $\delta_{m,k_x}$ is generically nonzero, and no single shift
$\varphi\mapsto\varphi+\varphi_0$ makes the complete channel current odd.
This harmonic phase unlocking is the origin of the finite single-channel
diode response.

For completeness, when the first two harmonics dominate, introducing the shifted phase
$\chi=\varphi+\arg\mathcal C_{1,k_x}$ gives
\begin{equation}
I_{k_x}(\chi)
=A_{1,k_x}\sin\chi
+A_{2,k_x}\sin(2\chi+\delta_{2,k_x}),
\qquad
\delta_{2,k_x}
=\arg\mathcal C_{2,k_x}-2\arg\mathcal C_{1,k_x}.
\label{eq:s-two-harmonic-channel}
\end{equation}
For $r_{k_x}=A_{2,k_x}/A_{1,k_x}\ll1$, the two critical currents are
\begin{align}
I_{c,k_x}^{+}
&=A_{1,k_x}-A_{2,k_x}\sin\delta_{2,k_x}
+\mathcal O(r_{k_x}^2A_{1,k_x}),\nonumber\\
I_{c,k_x}^{-}
&=-A_{1,k_x}-A_{2,k_x}\sin\delta_{2,k_x}
+\mathcal O(r_{k_x}^2A_{1,k_x}),
\label{eq:s-channel-critical-currents}
\end{align}
and hence
\begin{equation}
\eta_{k_x}
=-r_{k_x}\sin\delta_{2,k_x}
+\mathcal O(r_{k_x}^2).
\label{eq:s-small-channel-eta}
\end{equation}
An individual channel is therefore weakly rectifying when both its
higher-harmonic ratio $r_{k_x}$ and its phase mismatch
$\delta_{2,k_x}$ are small.  At the reference parameters,
$\max_{k_x}|\eta_{k_x}|=1.29\times10^{-4}$.

\begin{figure}[t]
\centering
\includegraphics[width=0.98\textwidth]{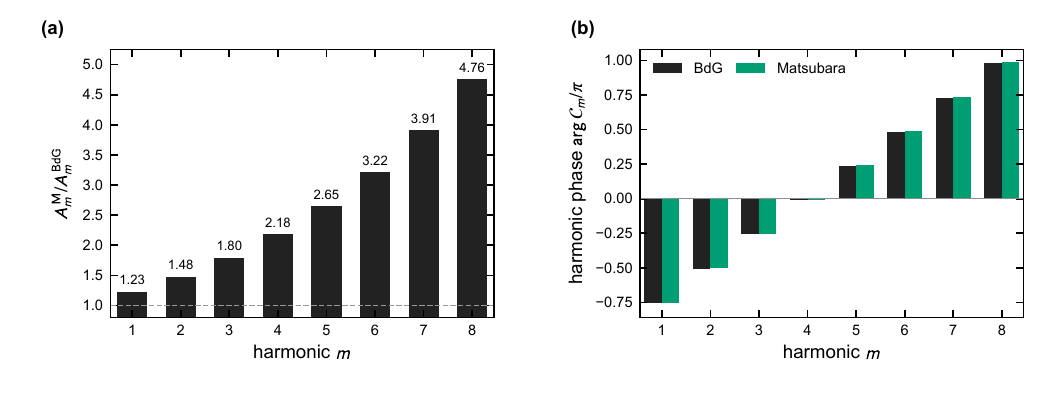}
\caption{\label{fig:s-harmonic-amplitude-phase}
Fixed-channel Fourier comparison between the Matsubara theory and the BdG
calculation at $k_x=0.3\pi$ and $k_BT=0.05\Delta_0$.
(a) Harmonic-amplitude ratio $A_m^{\mathrm M}/A_m^{\mathrm{BdG}}$ for
$m=1,\ldots,8$; the numbers above the bars give the corresponding ratios,
and the dashed line marks unity.
(b) Harmonic phases $\arg\mathcal C_m/\pi$ from BdG (black) and Matsubara
(green), shown on their common principal branch $(-\pi,\pi]$.  The remaining
parameters are the same as in Fig.~3 of the main text.}
\end{figure}

Figure~\ref{fig:s-harmonic-amplitude-phase} separates the fixed-channel
benchmark of Fig.~3(a,b) into amplitude and phase components.  For
$k_x=0.3\pi$, the ratio $A_m^{\mathrm M}/A_m^{\mathrm{BdG}}$ increases from
$1.23$ at $m=1$ to $4.76$ at $m=8$.  Panel (b) places the corresponding
harmonic phases on the same principal branch.  Their wrapped difference grows
from $9.97\times10^{-4}\pi$ at $m=1$ to $7.94\times10^{-3}\pi$ at $m=8$,
equivalent to an almost order-independent phase error
$|\Delta\phi_m|/m\simeq9.93\times10^{-4}\pi$.  The stable-zero difference is
$1.00\times10^{-3}\pi$.  Thus the scattering theory reproduces the phase
structure at the $10^{-3}\pi$-per-order scale while progressively
overestimating the higher-harmonic amplitudes.

Equation~(\ref{eq:s-matsubara-current}) is the Matsubara theory used in the
main text.  It retains the active bonding dispersion and its full
energy-dependent propagation phase, and therefore captures the channel
anomalous phase accurately and its harmonic phase unlocking
semiquantitatively.  Because it treats the
interfaces as transparent scalar Andreev reflectors, it does not include
ordinary interface reflection, spin--sublattice-dependent reflection, or
propagation through the finite superconducting slabs.  These effects
renormalize the harmonic amplitudes and are retained by the Green-function
treatment below.

\FloatBarrier
\section{Exact Green-function reduction and effective bonding theory}

This section integrates out the finite superconducting regions exactly and
tests whether the resulting Josephson current is dominated by the active
bonding-derived sector.

Introduce
\begin{equation}
\mathcal M(i\omega_n,\varphi)
=i\omega_n-\mathcal H_{\mathrm{SNS}}(\varphi)
\label{eq:s-M}
\end{equation}
and partition the superconducting and normal sites:
\begin{equation}
\mathcal M=
\begin{pmatrix}
\mathcal M_{SS}&\mathcal M_{SN}\\
\mathcal M_{NS}&\mathcal M_{NN}
\end{pmatrix}.
\label{eq:s-partition}
\end{equation}
The determinant identity for a block matrix gives
\begin{align}
\det\mathcal M
&=\det\mathcal M_{SS}\det\mathcal G_N^{-1},
\label{eq:s-determinant}\\
\mathcal G_N^{-1}
&=\mathcal M_{NN}
-\mathcal M_{NS}\mathcal M_{SS}^{-1}\mathcal M_{SN}\nonumber\\
&=i\omega_n-H_N-\Sigma_S(i\omega_n,\varphi).
\label{eq:s-normal-green}
\end{align}
Here
$\Sigma_S=\mathcal M_{NS}\mathcal M_{SS}^{-1}\mathcal M_{SN}$ is a matrix
self-energy.  Unlike a phenomenological transparency, it retains the full
frequency dependence, finite-slab spectrum, and spin--sublattice structure
of all normal and Andreev reflection processes at both interfaces.
After the normal sites are removed, the two superconducting slabs are
disconnected.  Their phases can be gauged away independently, so
$\det\mathcal M_{SS}$ is independent of $\varphi$.  The phase-dependent
thermodynamic potential and current are therefore
\begin{align}
\Omega(\varphi)
&=-\frac{1}{2\beta}
\sum_{\omega_n}\ln\det\mathcal G_N^{-1},
\label{eq:s-free-energy}\\
I(\varphi)
&=-\frac{e}{\hbar\beta}
\sum_{\omega_n}
\operatorname{Tr}\left[
\mathcal G_N\partial_{\varphi}\mathcal G_N^{-1}
\right].
\label{eq:s-green-current}
\end{align}
We next reduce the exact normal-region Green function to the active bonding
sector.  At fixed $k_x$, the electron-sector intersublattice hopping operator
restricted to the normal region is
\begin{equation}
 h_{\mathrm{inter}}^e(k_x)
 =
 \begin{pmatrix}
  0&T_{AB}(k_x)\\
  T_{AB}^\dagger(k_x)&0
 \end{pmatrix}
 \otimes\sigma_0 .
\label{eq:s-inter-hopping-operator}
\end{equation}
Here $T_{AB}(k_x)$ contains the four $A$--$B$ hopping amplitudes
$-t/2$, including the Bloch factors associated with their $x$-directed cell
displacements.  We diagonalize this Hermitian operator according to
\begin{equation}
 h_{\mathrm{inter}}^e(k_x)u_{\ell,\pm}(k_x)
 =
 \pm\epsilon_\ell(k_x)u_{\ell,\pm}(k_x),
 \qquad \epsilon_\ell(k_x)>0 .
\label{eq:s-inter-hopping-eigenvectors}
\end{equation}
The negative-eigenvalue states are the bonding states.  Thus
$U_b(k_x)=[u_{1,-}(k_x),\ldots,u_{N_b,-}(k_x)]$ is an isometry whose columns
form an orthonormal basis of the electron-sector bonding subspace.  At an
isolated momentum where the hopping splitting vanishes, this subspace is
defined by continuous continuation from neighboring momenta.  The associated
electron and BdG projectors are
\begin{align}
 P_b^e(k_x)&=U_b(k_x)U_b^\dagger(k_x),\nonumber\\
 P_b(k_x)&=\operatorname{diag}\!\left[
 P_b^e(k_x),P_b^{e*}(-k_x)
 \right],\qquad P_r=1-P_b.
\label{eq:s-bdg-projector}
\end{align}
The second line is the projector in the time-reversal-covariant Nambu basis.
The complementary projector $P_r$ refers to the antibonding sector.  With
$\mathcal M_{rb}=P_r\mathcal G_N^{-1}P_b$, the inverse Green function becomes
\begin{equation}
\mathcal G_N^{-1}=
\begin{pmatrix}
\mathcal M_{bb}&\mathcal M_{br}\\
\mathcal M_{rb}&\mathcal M_{rr}
\end{pmatrix}.
\label{eq:s-band-blocks}
\end{equation}
Integrating out the complementary sector produces
\begin{equation}
\mathcal M_b^{\mathrm{eff}}
=\mathcal M_{bb}
-\mathcal M_{br}\mathcal M_{rr}^{-1}\mathcal M_{rb}.
\label{eq:s-bonding-schur}
\end{equation}
The second term is the self-energy associated with virtual processes
$b\rightarrow r\rightarrow b$.  A direct projection retains only
$\mathcal M_{bb}$ and is exact only if
$[\mathcal H_{\mathrm{SNS}},P_b]=0$, which does not hold in the presence of
staggered spin-orbit coupling, pairing interfaces, and finite transverse
geometry.  The further determinant identity
\begin{equation}
 \det\mathcal G_N^{-1}
 =\det\mathcal M_{rr}\det\mathcal M_b^{\mathrm{eff}}
\label{eq:s-band-determinant}
\end{equation}
separates direct phase-dependent processes in the complementary sector from
its virtual dressing of the bonding sector.
\begin{figure}[!htbp]
\centering
\includegraphics[width=0.98\textwidth]
{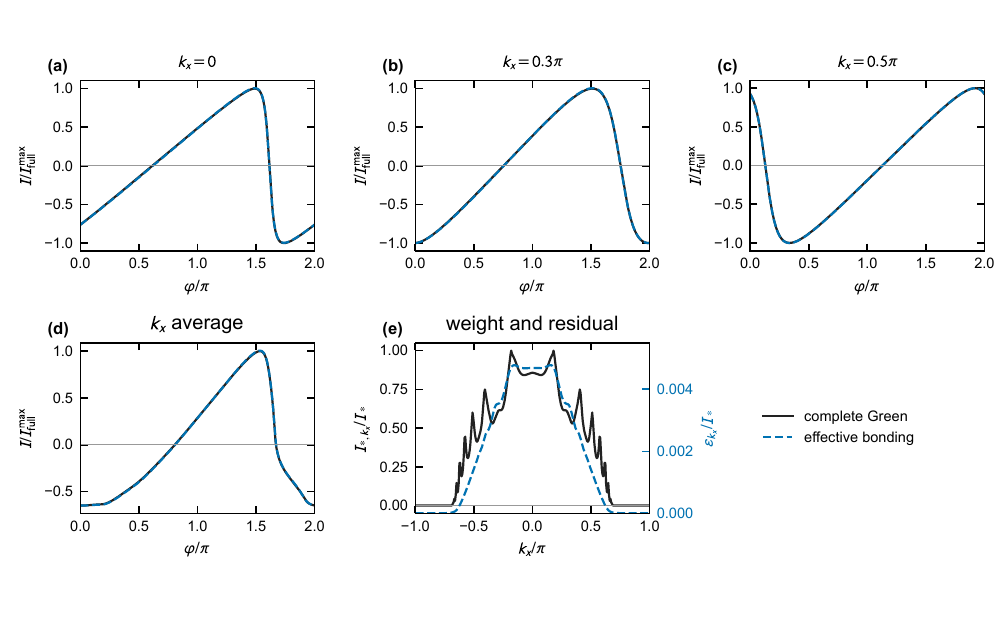}
\caption{\label{fig:s-multikx-validation}
Validation of the effective bonding-sector description at
$\mu=-1.1t$, $J_n=0.4t$, and $\lambda_N=0.25t$, with $t'=0.1t$, $\Delta_0=0.075t$, $d_N=12$, $W_S=20$, $J_S=\lambda_S=0$, and zero interface SOC.
(a)--(c) Complete Green-function CPR (solid black) and the
Schur-complement bonding result (dashed blue) for three representative
transverse momenta.  Each panel is normalized by the maximum magnitude of
its complete current.
(d) The corresponding currents after summing $N_k=321$ midpoint-sampled
transverse channels.
(e) Normalized channel current scale $I_{*,k_x}/I_*$ (black) and normalized
Schur residual $\epsilon_{k_x}/I_*$ (blue).  The calculation uses
$N_\varphi=1441$ phase points and $N_\omega=384$ mapped Gauss--Legendre points
for the zero-temperature Matsubara integral.}
\end{figure}

The validity of this reduction is tested beyond a single representative
channel in Fig.~\ref{fig:s-multikx-validation}.  Panels (a)--(c) compare the
complete Green-function CPR with the Schur-complement bonding result at
$k_x=0$, $0.3\pi$, and $0.5\pi$.  Their maximum deviations, normalized by the
maximum current of the corresponding complete channel, are $0.547\%$,
$0.586\%$, and $0.270\%$, respectively.  Panel (d) performs the physically
relevant sum over $N_k=321$ transverse momenta before comparing the two
currents.  The maximum deviation is then $0.559\%$, while the full and
effective-bonding currents give
\begin{equation}
 \eta_{\mathrm{full}}=0.21241,\qquad
 \eta_b^{\mathrm{eff}}=0.21473.
\label{eq:s-multikx-eta}
\end{equation}
As an independent check, at $k_x=0.3\pi$ the complete Green-function current
agrees with direct diagonalization of the full BdG Hamiltonian to a maximum
absolute difference of $4.8\times10^{-14}$ on the same phase grid.

To display where the comparison is relevant,
Fig.~\ref{fig:s-multikx-validation}(e) shows the channel current scale
$I_{*,k_x}=\max_\varphi|I_{\mathrm{full},k_x}(\varphi)|$ together with the
absolute Schur residual
\begin{equation}
 \epsilon_{k_x}
 =\max_\varphi\left|
 I_{\mathrm{full},k_x}(\varphi)
 -I_{b,k_x}^{\mathrm{eff}}(\varphi)
 \right|.
\label{eq:s-schur-residual}
\end{equation}
Both are normalized by $I_*=\max_{k_x}I_{*,k_x}$.  The residual remains below
$0.0048I_*$ throughout the Brillouin zone.  By
Eq.~(\ref{eq:s-band-determinant}), this residual is the direct
phase-dependent current carried by the complementary determinant, whereas
its indirect effect is retained exactly in the Schur self-energy of
Eq.~(\ref{eq:s-bonding-schur}).  The comparison therefore supports an
effective single-active-band description for the reference parameter set:
the $\mathcal{PT}$-degenerate bonding-derived branch carries the dominant
Josephson current, while the complementary sector mainly dresses its
frequency-dependent effective Green function.

\FloatBarrier
\section{Magnetic-point-group classification and materials}
\label{sec:pt-materials}

This section specifies the directional symmetry criterion and lists the
associated materials by magnetic point group. We consider the 21 groups
with $\mathcal{PT}\in G$ and $\mathcal P,\mathcal T\notin G$, rather than
all magnetic groups that contain $\mathcal{PT}$.
For $G=H\cup(\mathcal{PT})H$, $H$ denotes the unitary subgroup.
A unitary operation with spatial matrix $R$ maps $\mathbf k$ to
$R\mathbf k$, whereas $R\mathcal T$ maps it to $-R\mathbf k$.
Because $\mathcal{PT}$ leaves momentum unchanged, all momentum mappings
generated by $G$ are already represented by $H$.
For a unit direction vector $\hat{\mathbf u}$, an operation
$h\in H$ with $h\hat{\mathbf u}=-\hat{\mathbf u}$ therefore enforces
equality of the spectra at $q\hat{\mathbf u}$ and
$-q\hat{\mathbf u}$, allowing a permutation of band labels.
In the absence of such an operation, band nonreciprocity along this
direction is symmetry allowed. The test concerns directions through
$\Gamma$; momenta at special Brillouin-zone boundaries must also be
identified modulo reciprocal lattice vectors.

In the direction table in the main text, the monoclinic twofold axis or
mirror normal is $z$, and the primed mirror of $m'mm$ is normal to $x$.
The tetragonal, trigonal, and hexagonal principal axes are $z$.
For $H=32,422,622$, one unitary twofold axis is chosen along $x$;
for $H=3m$ and $\bar6m2$, a unitary vertical mirror is normal to $x$.
The other tetragonal and hexagonal groups use the conventional basal
axes, and cubic groups use conventional cubic axes.
These are point-group coordinates, not automatically the crystallographic
$a,b,c$ axes of every material. In particular, three forbidden coordinate
axes do not exclude allowed generic three-dimensional directions, whereas
a unitary twofold rotation normal to a chosen two-dimensional plane
enforces reciprocal spectra throughout that plane.

Table~\ref{tab:pt-materials} reorganizes the magnetic-structure records
in Supplemental Table XI of Ref.~\cite{Xiao2023MagneticSHG} according to
this point-group classification. The selection contains 269 records
covering 15 group types. Repeated chemical formulas within the same group
are combined into 217 material--group entries, while every MAGNDATA
BCS-ID is retained; the same formula may consequently appear under
different groups. The MAGNDATA ID identifies a magnetic structure.
The $x$, $y$, and $z$ columns indicate symmetry-allowed band nonreciprocity
in the point-group coordinates specified above, rather than the
crystallographic axes of each individual material.
The groups $6/m'm'm'$, $6/m'mm$, $6'/mmm'$, $m'\bar3'$,
$m'\bar3'm$, and $m'\bar3'm'$ have no entries in this source selection;
this does not imply that they lack material realizations.

The group assignments refer to the recorded magnetic phases, including
their crystal structures and magnetic configurations, and should not be
treated as permanent labels of chemical formulas. Axis-permuted symbols
such as $m'mm$, $mm'm$, and $mmm'$ are grouped into the same type, but
their transport directions must be transformed with the axes.
For thin films or few-layer samples, the magnetic domain, stacking,
surface, and contact geometry must be specified before applying the
criterion to a Josephson junction. The catalogue is not restricted to
metallic weak links or to the two-sublattice model studied in the main
text, and symmetry permission alone does not establish a finite diode
efficiency.

\begingroup
\small
\setlength{\tabcolsep}{4pt}
\renewcommand{\arraystretch}{1.06}
\setlength{\LTleft}{0pt}
\setlength{\LTright}{0pt}
\setlength{\LTcapwidth}{\textwidth}
\setlength{\LTpre}{\medskipamount}
\setlength{\LTpost}{\medskipamount}
\setlength{\SMwidecol}{\dimexpr(\textwidth-8\tabcolsep)*4/11\relax}
\setlength{\SMnarrowcol}{\dimexpr(\textwidth-8\tabcolsep)/11\relax}
\begin{longtable}{@{}llccc@{}}
\caption{Materials grouped by magnetic point group, reorganized from
Supplemental Table XI of Ref.~\cite{Xiao2023MagneticSHG}.
All 269 source MAGNDATA IDs are retained; multiple IDs in a row denote
separate magnetic-structure records with the same chemical formula and
group type. In the point-group coordinates specified in this section,
$\checkmark$ denotes symmetry-allowed band nonreciprocity along the
indicated axis and $\times$ denotes symmetry-enforced reciprocity.
These axes are not automatically the crystallographic axes of each
material, and three $\times$ symbols do not exclude nonreciprocity along
generic oblique directions.}\label{tab:pt-materials}\\
\hline
\SMmaterialcell{Material} & \SMidcell{MAGNDATA ID} &
\makebox[\SMnarrowcol][c]{$x$} & \makebox[\SMnarrowcol][c]{$y$} &
\makebox[\SMnarrowcol][c]{$z$} \\
\hline
\endfirsthead
\multicolumn{5}{c}{\normalfont\small\tablename~\thetable~(continued)}\\
\hline
\SMmaterialcell{Material} & \SMidcell{MAGNDATA ID} &
\makebox[\SMnarrowcol][c]{$x$} & \makebox[\SMnarrowcol][c]{$y$} &
\makebox[\SMnarrowcol][c]{$z$} \\
\hline
\endhead
\hline
\multicolumn{5}{r}{\normalfont\small Continued on next page}\\
\endfoot
\hline
\endlastfoot
\multicolumn{5}{@{}l@{}}{\textbf{Magnetic point group }$\bar{1}'$ \qquad $H=1$ (7 records)}\\*\hline\nopagebreak
\SMmaterialcell{$\mathrm{CaMnGe_{2}O_{6}}$} & \SMidcell{0.155} & $\checkmark$ & $\checkmark$ & $\checkmark$ \\
\SMmaterialcell{$\mathrm{MnPSe_{3}}$} & \SMidcell{0.180,\allowbreak 0.524} & $\checkmark$ & $\checkmark$ & $\checkmark$ \\
\SMmaterialcell{$\mathrm{BaNi_{2}P_{2}O_{8}}$} & \SMidcell{0.215} & $\checkmark$ & $\checkmark$ & $\checkmark$ \\
\SMmaterialcell{$\mathrm{YbMn_{2}Sb_{2}}$} & \SMidcell{0.483} & $\checkmark$ & $\checkmark$ & $\checkmark$ \\
\SMmaterialcell{$\mathrm{NaCrSi_{2}O_{6}}$} & \SMidcell{0.504} & $\checkmark$ & $\checkmark$ & $\checkmark$ \\
\SMmaterialcell{$\mathrm{CaMn_{2}Sb_{2}}$} & \SMidcell{0.523} & $\checkmark$ & $\checkmark$ & $\checkmark$ \\
\hline
\multicolumn{5}{@{}l@{}}{\textbf{Magnetic point group }$2'/m$ \qquad $H=m_{z}$ (29 records)}\\*\hline\nopagebreak
\SMmaterialcell{$\mathrm{CaMn_{2}Sb_{2}}$} & \SMidcell{0.92} & $\checkmark$ & $\checkmark$ & $\times$ \\
\SMmaterialcell{$\mathrm{Cr_{2}O_{3}}$} & \SMidcell{0.110} & $\checkmark$ & $\checkmark$ & $\times$ \\
\SMmaterialcell{$\mathrm{Co_{3}TeO_{6}}$} & \SMidcell{0.145} & $\checkmark$ & $\checkmark$ & $\times$ \\
\SMmaterialcell{$\mathrm{CaMnGe_{2}O_{6}}$} & \SMidcell{0.156} & $\checkmark$ & $\checkmark$ & $\times$ \\
\SMmaterialcell{$\mathrm{MnPS_{3}}$} & \SMidcell{0.163} & $\checkmark$ & $\checkmark$ & $\times$ \\
\SMmaterialcell{$\mathrm{CeMnAsO}$} & \SMidcell{0.188} & $\checkmark$ & $\checkmark$ & $\times$ \\
\SMmaterialcell{$\mathrm{TlFe_{1.6}Se_{2}}$} & \SMidcell{0.208} & $\checkmark$ & $\checkmark$ & $\times$ \\
\SMmaterialcell{$\mathrm{LiCrGe_{2}O_{6}}$} & \SMidcell{0.217} & $\checkmark$ & $\checkmark$ & $\times$ \\
\SMmaterialcell{$\mathrm{Li_{2}Fe(SO_{4})_{2}}$} & \SMidcell{0.243} & $\checkmark$ & $\checkmark$ & $\times$ \\
\SMmaterialcell{$\mathrm{Li_{1.5}Fe(SO_{4})_{2}}$} & \SMidcell{0.245} & $\checkmark$ & $\checkmark$ & $\times$ \\
\SMmaterialcell{$\mathrm{Cs_{2}FeCl_{5}.D_{2}O}$} & \SMidcell{0.252} & $\checkmark$ & $\checkmark$ & $\times$ \\
\SMmaterialcell{$\mathrm{MnGeO_{3}}$} & \SMidcell{0.312} & $\checkmark$ & $\checkmark$ & $\times$ \\
\SMmaterialcell{$\mathrm{Er_{2}ReC_{2}}$} & \SMidcell{0.347} & $\checkmark$ & $\checkmark$ & $\times$ \\
\SMmaterialcell{$\mathrm{DyCrO_{4}}$} & \SMidcell{0.372} & $\checkmark$ & $\checkmark$ & $\times$ \\
\SMmaterialcell{$\mathrm{LiCoPO_{4}}$} & \SMidcell{0.384} & $\checkmark$ & $\checkmark$ & $\times$ \\
\SMmaterialcell{$\mathrm{YbCl_{3}}$} & \SMidcell{0.444,\allowbreak 0.585,\allowbreak 0.723} & $\checkmark$ & $\checkmark$ & $\times$ \\
\SMmaterialcell{$\mathrm{Cs_{2}[FeCl_{5}(H_{2}O)]}$} & \SMidcell{0.476} & $\checkmark$ & $\checkmark$ & $\times$ \\
\SMmaterialcell{$\mathrm{SrMn_{2}As_{2}}$} & \SMidcell{0.482} & $\checkmark$ & $\checkmark$ & $\times$ \\
\SMmaterialcell{$\mathrm{NdB_{4}}$} & \SMidcell{0.492} & $\checkmark$ & $\checkmark$ & $\times$ \\
\SMmaterialcell{$\mathrm{Co_{4}Ta_{2}O_{9}}$} & \SMidcell{0.511} & $\checkmark$ & $\checkmark$ & $\times$ \\
\SMmaterialcell{$\mathrm{Er_{2}Si_{2}O_{7}}$} & \SMidcell{0.527} & $\checkmark$ & $\checkmark$ & $\times$ \\
\SMmaterialcell{$\mathrm{KFeS_{2}}$} & \SMidcell{0.633} & $\checkmark$ & $\checkmark$ & $\times$ \\
\SMmaterialcell{$\mathrm{RbFeS_{2}}$} & \SMidcell{0.636} & $\checkmark$ & $\checkmark$ & $\times$ \\
\SMmaterialcell{$\mathrm{ErSi_{2}O_{7}}$} & \SMidcell{0.650} & $\checkmark$ & $\checkmark$ & $\times$ \\
\SMmaterialcell{$\mathrm{Mn_{3}Ta_{2}O_{8}}$} & \SMidcell{0.734} & $\checkmark$ & $\checkmark$ & $\times$ \\
\SMmaterialcell{$\mathrm{Ag_{2}CrO_{2}}$} & \SMidcell{1.0.1} & $\checkmark$ & $\checkmark$ & $\times$ \\
\SMmaterialcell{$\mathrm{HoBaCuO_{5}}$} & \SMidcell{2.85} & $\checkmark$ & $\checkmark$ & $\times$ \\
\hline
\multicolumn{5}{@{}l@{}}{\textbf{Magnetic point group }$2/m'$ \qquad $H=2_{z}$ (29 records)}\\*\hline\nopagebreak
\SMmaterialcell{$\mathrm{LiFeSi_{2}O_{6}}$} & \SMidcell{0.28} & $\times$ & $\times$ & $\checkmark$ \\
\SMmaterialcell{$\mathrm{LiFePO_{4}}$} & \SMidcell{0.152} & $\times$ & $\times$ & $\checkmark$ \\
\SMmaterialcell{$\mathrm{Co_{4}Nb_{2}O_{9}}$} & \SMidcell{0.196,\allowbreak 0.197,\allowbreak 0.529} & $\times$ & $\times$ & $\checkmark$ \\
\SMmaterialcell{$\mathrm{Fe_{3}(PO_{4})_{2}}$} & \SMidcell{0.264} & $\times$ & $\times$ & $\checkmark$ \\
\SMmaterialcell{$\mathrm{Co_{2}V_{2}O_{7}}$} & \SMidcell{0.281} & $\times$ & $\times$ & $\checkmark$ \\
\SMmaterialcell{$\mathrm{ErGe_{3}}$} & \SMidcell{0.330} & $\times$ & $\times$ & $\checkmark$ \\
\SMmaterialcell{$\mathrm{LiCoPO_{4}}$} & \SMidcell{0.385} & $\times$ & $\times$ & $\checkmark$ \\
\SMmaterialcell{$\mathrm{Cu_{2}CdB_{2}O_{6}}$} & \SMidcell{0.394} & $\times$ & $\times$ & $\checkmark$ \\
\SMmaterialcell{$\mathrm{EuMnSb_{2}}$} & \SMidcell{0.422} & $\times$ & $\times$ & $\checkmark$ \\
\SMmaterialcell{$\mathrm{Fe_{4}Nb_{2}O_{9}}$} & \SMidcell{0.441,\allowbreak 0.442,\allowbreak 0.443} & $\times$ & $\times$ & $\checkmark$ \\
\SMmaterialcell{$\mathrm{Pb_{2}VO(PO_{4})_{2}}$} & \SMidcell{0.505} & $\times$ & $\times$ & $\checkmark$ \\
\SMmaterialcell{$\mathrm{CaMnGe}$} & \SMidcell{0.601,\allowbreak 0.602} & $\times$ & $\times$ & $\checkmark$ \\
\SMmaterialcell{$\mathrm{KFeSe_{2}}$} & \SMidcell{0.637} & $\times$ & $\times$ & $\checkmark$ \\
\SMmaterialcell{$\mathrm{RbFeSe_{2}}$} & \SMidcell{0.638} & $\times$ & $\times$ & $\checkmark$ \\
\SMmaterialcell{$\mathrm{MoP_{3}SiO_{11}}$} & \SMidcell{0.728,\allowbreak 0.804} & $\times$ & $\times$ & $\checkmark$ \\
\SMmaterialcell{$\mathrm{Fe_{2}Co_{2}Nb_{2}O_{9}}$} & \SMidcell{0.770} & $\times$ & $\times$ & $\checkmark$ \\
\SMmaterialcell{$\mathrm{Fe_{2}WO_{6}}$} & \SMidcell{0.809} & $\times$ & $\times$ & $\checkmark$ \\
\SMmaterialcell{$\mathrm{Na_{2}MnPO_{4}F}$} & \SMidcell{0.827,\allowbreak 0.828,\allowbreak 0.829,\allowbreak 0.830} & $\times$ & $\times$ & $\checkmark$ \\
\SMmaterialcell{$\mathrm{SrHo_{2}O_{4}}$} & \SMidcell{2.8} & $\times$ & $\times$ & $\checkmark$ \\
\SMmaterialcell{$\mathrm{TbOOH}$} & \SMidcell{2.21} & $\times$ & $\times$ & $\checkmark$ \\
\hline
\multicolumn{5}{@{}l@{}}{\textbf{Magnetic point group }$m'm'm'$ \qquad $H=222$ (26 records)}\\*\hline\nopagebreak
\SMmaterialcell{$\mathrm{LiMnPO_{4}}$} & \SMidcell{0.24,\allowbreak 0.382} & $\times$ & $\times$ & $\times$ \\
\SMmaterialcell{$\mathrm{YFe_{4}Ge_{2}}$} & \SMidcell{0.27} & $\times$ & $\times$ & $\times$ \\
\SMmaterialcell{$\mathrm{Li_{2}Ni(SO_{4})_{2}}$} & \SMidcell{0.71} & $\times$ & $\times$ & $\times$ \\
\SMmaterialcell{$\mathrm{LuFe_{4}Ge_{2}}$} & \SMidcell{0.140} & $\times$ & $\times$ & $\times$ \\
\SMmaterialcell{$\mathrm{EuZrO_{3}}$} & \SMidcell{0.147} & $\times$ & $\times$ & $\times$ \\
\SMmaterialcell{$\mathrm{DyCoO_{3}}$} & \SMidcell{0.159,\allowbreak 0.521} & $\times$ & $\times$ & $\times$ \\
\SMmaterialcell{$\mathrm{DyScO_{3}}$} & \SMidcell{0.171} & $\times$ & $\times$ & $\times$ \\
\SMmaterialcell{$\mathrm{Li_{2}Co(SO_{4})_{2}}$} & \SMidcell{0.244} & $\times$ & $\times$ & $\times$ \\
\SMmaterialcell{$\mathrm{LiFe(SO_{4})_{2}}$} & \SMidcell{0.246} & $\times$ & $\times$ & $\times$ \\
\SMmaterialcell{$\mathrm{CeCu_{2}}$} & \SMidcell{0.290} & $\times$ & $\times$ & $\times$ \\
\SMmaterialcell{$\mathrm{TbAlO_{3}}$} & \SMidcell{0.350} & $\times$ & $\times$ & $\times$ \\
\SMmaterialcell{$\mathrm{RbFeCl_{5}(D_{2}O)}$} & \SMidcell{0.362} & $\times$ & $\times$ & $\times$ \\
\SMmaterialcell{$\mathrm{KFeCl_{5}(D_{2}O)}$} & \SMidcell{0.363} & $\times$ & $\times$ & $\times$ \\
\SMmaterialcell{$\mathrm{Sr_{2}Fe_{1.9}Co_{0.1}O_{5.5}}$} & \SMidcell{0.400} & $\times$ & $\times$ & $\times$ \\
\SMmaterialcell{$\mathrm{Sr_{4}Fe_{4}O_{11}}$} & \SMidcell{0.401} & $\times$ & $\times$ & $\times$ \\
\SMmaterialcell{$\mathrm{GdAlO_{3}}$} & \SMidcell{0.410} & $\times$ & $\times$ & $\times$ \\
\SMmaterialcell{$\mathrm{EuMnSb_{2}}$} & \SMidcell{0.421,\allowbreak 0.423,\allowbreak 0.424} & $\times$ & $\times$ & $\times$ \\
\SMmaterialcell{$\mathrm{RbFeO_{2}}$} & \SMidcell{0.455} & $\times$ & $\times$ & $\times$ \\
\SMmaterialcell{$\mathrm{CsFeO_{2}}$} & \SMidcell{0.457} & $\times$ & $\times$ & $\times$ \\
\SMmaterialcell{$\mathrm{TbB_{4}}$} & \SMidcell{0.469} & $\times$ & $\times$ & $\times$ \\
\SMmaterialcell{$\mathrm{SrFe_{2}Se_{2}O}$} & \SMidcell{0.761} & $\times$ & $\times$ & $\times$ \\
\SMmaterialcell{$\mathrm{SrFe_{2}S_{2}O}$} & \SMidcell{0.762} & $\times$ & $\times$ & $\times$ \\
\hline
\multicolumn{5}{@{}l@{}}{\textbf{Magnetic point group }$m'mm$ \qquad $H=mm2_{(x)}$ (78 records)}\\*\hline\nopagebreak
\SMmaterialcell{$\mathrm{U_{3}Ru_{4}Al_{12}}$} & \SMidcell{0.12} & $\checkmark$ & $\times$ & $\times$ \\
\SMmaterialcell{$\mathrm{Gd_{5}Ge_{4}}$} & \SMidcell{0.14} & $\checkmark$ & $\times$ & $\times$ \\
\SMmaterialcell{$\mathrm{EuTiO_{3}}$} & \SMidcell{0.16} & $\checkmark$ & $\times$ & $\times$ \\
\SMmaterialcell{$\mathrm{DyB_{4}}$} & \SMidcell{0.22} & $\checkmark$ & $\times$ & $\times$ \\
\SMmaterialcell{$\mathrm{Cr_{2}WO_{6}}$} & \SMidcell{0.75,\allowbreak 0.144} & $\checkmark$ & $\times$ & $\times$ \\
\SMmaterialcell{$\mathrm{Cr_{2}TeO_{6}}$} & \SMidcell{0.76,\allowbreak 0.143} & $\checkmark$ & $\times$ & $\times$ \\
\SMmaterialcell{$\mathrm{KMn_{4}(PO_{4})_{3}}$} & \SMidcell{0.86} & $\checkmark$ & $\times$ & $\times$ \\
\SMmaterialcell{$\mathrm{NaFePO_{4}}$} & \SMidcell{0.87} & $\checkmark$ & $\times$ & $\times$ \\
\SMmaterialcell{$\mathrm{LiNiPO_{4}}$} & \SMidcell{0.88} & $\checkmark$ & $\times$ & $\times$ \\
\SMmaterialcell{$\mathrm{LiFePO_{4}}$} & \SMidcell{0.95} & $\checkmark$ & $\times$ & $\times$ \\
\SMmaterialcell{$\mathrm{CoSe_{2}O_{5}}$} & \SMidcell{0.119,\allowbreak 0.161} & $\checkmark$ & $\times$ & $\times$ \\
\SMmaterialcell{$\mathrm{Tb_{5}Ge_{4}}$} & \SMidcell{0.141,\allowbreak 0.411,\allowbreak 0.412} & $\checkmark$ & $\times$ & $\times$ \\
\SMmaterialcell{$\mathrm{EuZrO_{3}}$} & \SMidcell{0.146} & $\checkmark$ & $\times$ & $\times$ \\
\SMmaterialcell{$\mathrm{TbCoO_{3}}$} & \SMidcell{0.160,\allowbreak 0.520} & $\checkmark$ & $\times$ & $\times$ \\
\SMmaterialcell{$\mathrm{NdCrTiO_{5}}$} & \SMidcell{0.162} & $\checkmark$ & $\times$ & $\times$ \\
\SMmaterialcell{$\mathrm{KCrF_{4}}$} & \SMidcell{0.182} & $\checkmark$ & $\times$ & $\times$ \\
\SMmaterialcell{$\mathrm{CeMnAsO}$} & \SMidcell{0.187} & $\checkmark$ & $\times$ & $\times$ \\
\SMmaterialcell{$\mathrm{LiCoPO_{4}}$} & \SMidcell{0.193,\allowbreak 0.383} & $\checkmark$ & $\times$ & $\times$ \\
\SMmaterialcell{$\mathrm{SrEr_{2}O_{4}}$} & \SMidcell{0.216} & $\checkmark$ & $\times$ & $\times$ \\
\SMmaterialcell{$\mathrm{CuMnAs}$} & \SMidcell{0.222} & $\checkmark$ & $\times$ & $\times$ \\
\SMmaterialcell{$\mathrm{Cu_{0.95}MnAs}$} & \SMidcell{0.223} & $\checkmark$ & $\times$ & $\times$ \\
\SMmaterialcell{$\mathrm{K_{2}CoP_{2}O_{7}}$} & \SMidcell{0.230} & $\checkmark$ & $\times$ & $\times$ \\
\SMmaterialcell{$\mathrm{CoGeO_{3}}$} & \SMidcell{0.311} & $\checkmark$ & $\times$ & $\times$ \\
\SMmaterialcell{$\mathrm{MnGeO_{3}}$} & \SMidcell{0.313} & $\checkmark$ & $\times$ & $\times$ \\
\SMmaterialcell{$\mathrm{DyGe_{1.75}}$} & \SMidcell{0.341} & $\checkmark$ & $\times$ & $\times$ \\
\SMmaterialcell{$\mathrm{TbGe_{2}}$} & \SMidcell{0.343} & $\checkmark$ & $\times$ & $\times$ \\
\SMmaterialcell{$\mathrm{Tb_{2}ReC_{2}}$} & \SMidcell{0.346} & $\checkmark$ & $\times$ & $\times$ \\
\SMmaterialcell{$\mathrm{Fe_{3}BO_{5}}$} & \SMidcell{0.386} & $\checkmark$ & $\times$ & $\times$ \\
\SMmaterialcell{$\mathrm{FeOOH}$} & \SMidcell{0.399} & $\checkmark$ & $\times$ & $\times$ \\
\SMmaterialcell{$\mathrm{GdNiSi_{3}}$} & \SMidcell{0.406} & $\checkmark$ & $\times$ & $\times$ \\
\SMmaterialcell{$\mathrm{CaCr_{0.86}Fe_{3.14}As_{3}}$} & \SMidcell{0.429} & $\checkmark$ & $\times$ & $\times$ \\
\SMmaterialcell{$\mathrm{DyRuAsO}$} & \SMidcell{0.451} & $\checkmark$ & $\times$ & $\times$ \\
\SMmaterialcell{$\mathrm{TbRuAsO}$} & \SMidcell{0.452} & $\checkmark$ & $\times$ & $\times$ \\
\SMmaterialcell{$\mathrm{DyCoSi_{2}}$} & \SMidcell{0.453} & $\checkmark$ & $\times$ & $\times$ \\
\SMmaterialcell{$\mathrm{KFeO_{2}}$} & \SMidcell{0.459,\allowbreak 0.460} & $\checkmark$ & $\times$ & $\times$ \\
\SMmaterialcell{$\mathrm{ThCr_{2}Si_{2}}$} & \SMidcell{0.466} & $\checkmark$ & $\times$ & $\times$ \\
\SMmaterialcell{$\mathrm{ErB_{4}}$} & \SMidcell{0.468} & $\checkmark$ & $\times$ & $\times$ \\
\SMmaterialcell{$\mathrm{TbNiGe_{2}}$} & \SMidcell{0.566} & $\checkmark$ & $\times$ & $\times$ \\
\SMmaterialcell{$\mathrm{HoNi_{0.64}Ge_{2}}$} & \SMidcell{0.567} & $\checkmark$ & $\times$ & $\times$ \\
\SMmaterialcell{$\mathrm{TbNi_{0.4}Ge_{2}}$} & \SMidcell{0.568} & $\checkmark$ & $\times$ & $\times$ \\
\SMmaterialcell{$\mathrm{TbCu_{0.4}Ge_{2}}$} & \SMidcell{0.569} & $\checkmark$ & $\times$ & $\times$ \\
\SMmaterialcell{$\mathrm{NdMnAsO}$} & \SMidcell{0.621,\allowbreak 0.622} & $\checkmark$ & $\times$ & $\times$ \\
\SMmaterialcell{$\mathrm{Mn_{2}Au}$} & \SMidcell{0.639,\allowbreak 0.640} & $\checkmark$ & $\times$ & $\times$ \\
\SMmaterialcell{$\mathrm{CeMnSbO}$} & \SMidcell{0.666} & $\checkmark$ & $\times$ & $\times$ \\
\SMmaterialcell{$\mathrm{PrMnSbO}$} & \SMidcell{0.668} & $\checkmark$ & $\times$ & $\times$ \\
\SMmaterialcell{$\mathrm{Ba_{4}Ru_{3}O_{10}}$} & \SMidcell{0.692,\allowbreak 0.693} & $\checkmark$ & $\times$ & $\times$ \\
\SMmaterialcell{$\mathrm{Bi_{2}CuO_{4}}$} & \SMidcell{0.695} & $\checkmark$ & $\times$ & $\times$ \\
\SMmaterialcell{$\mathrm{NdScO_{3}}$} & \SMidcell{0.782} & $\checkmark$ & $\times$ & $\times$ \\
\SMmaterialcell{$\mathrm{NdInO_{3}}$} & \SMidcell{0.783} & $\checkmark$ & $\times$ & $\times$ \\
\SMmaterialcell{$\mathrm{MnPd_{2}}$} & \SMidcell{0.798} & $\checkmark$ & $\times$ & $\times$ \\
\SMmaterialcell{$\mathrm{Tl_{3}Fe_{2}S_{4}}$} & \SMidcell{0.801} & $\checkmark$ & $\times$ & $\times$ \\
\SMmaterialcell{$\mathrm{DyBaCuO_{5}}$} & \SMidcell{0.805} & $\checkmark$ & $\times$ & $\times$ \\
\SMmaterialcell{$\mathrm{Fe_{2}Se_{2}O_{7}}$} & \SMidcell{0.806,\allowbreak 0.807,\allowbreak 0.808} & $\checkmark$ & $\times$ & $\times$ \\
\SMmaterialcell{$\mathrm{Fe_{2}WO_{6}}$} & \SMidcell{0.814} & $\checkmark$ & $\times$ & $\times$ \\
\SMmaterialcell{$\mathrm{MnNb_{2}O_{6}}$} & \SMidcell{0.815,\allowbreak 0.819} & $\checkmark$ & $\times$ & $\times$ \\
\SMmaterialcell{$\mathrm{MnTa_{2}O_{6}}$} & \SMidcell{0.816,\allowbreak 0.818} & $\checkmark$ & $\times$ & $\times$ \\
\SMmaterialcell{$\mathrm{Mn(Nb_{0.5}Ta_{0.5})_{2}O_{6}}$} & \SMidcell{0.817} & $\checkmark$ & $\times$ & $\times$ \\
\SMmaterialcell{$\mathrm{SrGd_{2}O_{4}}$} & \SMidcell{0.821} & $\checkmark$ & $\times$ & $\times$ \\
\SMmaterialcell{$\mathrm{Sr_{2}Mn_{3}Sb_{2}O_{2}}$} & \SMidcell{2.27} & $\checkmark$ & $\times$ & $\times$ \\
\SMmaterialcell{$\mathrm{Ba_{2}Mn_{3}Sb_{2}O_{2}}$} & \SMidcell{2.53} & $\checkmark$ & $\times$ & $\times$ \\
\SMmaterialcell{$\mathrm{La_{0.73}Tb_{0.27}Mn_{2}Si_{2}}$} & \SMidcell{2.58} & $\checkmark$ & $\times$ & $\times$ \\
\SMmaterialcell{$\mathrm{FeSn_{2}}$} & \SMidcell{2.66} & $\checkmark$ & $\times$ & $\times$ \\
\SMmaterialcell{$\mathrm{FeGe_{2}}$} & \SMidcell{2.68} & $\checkmark$ & $\times$ & $\times$ \\
\hline
\multicolumn{5}{@{}l@{}}{\textbf{Magnetic point group }$4/m'$ \qquad $H=4$ (5 records)}\\*\hline\nopagebreak
\SMmaterialcell{$\mathrm{Rb_{y}Fe_{2-x}Se_{2}}$} & \SMidcell{0.54} & $\times$ & $\times$ & $\checkmark$ \\
\SMmaterialcell{$\mathrm{K_{y}Fe_{2-x}Se_{2}}$} & \SMidcell{0.55} & $\times$ & $\times$ & $\checkmark$ \\
\SMmaterialcell{$\mathrm{TlFe_{1.6}Se_{2}}$} & \SMidcell{0.209} & $\times$ & $\times$ & $\checkmark$ \\
\SMmaterialcell{$\mathrm{K_{0.8}Fe_{1.8}Se_{2}}$} & \SMidcell{0.418} & $\times$ & $\times$ & $\checkmark$ \\
\SMmaterialcell{$\mathrm{NdB_{4}}$} & \SMidcell{0.491} & $\times$ & $\times$ & $\checkmark$ \\
\hline
\multicolumn{5}{@{}l@{}}{\textbf{Magnetic point group }$4'/m'$ \qquad $H=\bar{4}$ (2 records)}\\*\hline\nopagebreak
\SMmaterialcell{$\mathrm{KOsO_{4}}$} & \SMidcell{0.284} & $\times$ & $\times$ & $\times$ \\
\SMmaterialcell{$\mathrm{KRuO_{4}}$} & \SMidcell{0.285} & $\times$ & $\times$ & $\times$ \\
\hline
\multicolumn{5}{@{}l@{}}{\textbf{Magnetic point group }$4/m'm'm'$ \qquad $H=422$ (7 records)}\\*\hline\nopagebreak
\SMmaterialcell{$\mathrm{GdB_{4}}$} & \SMidcell{0.9} & $\times$ & $\times$ & $\times$ \\
\SMmaterialcell{$\mathrm{Fe_{2}TeO_{6}}$} & \SMidcell{0.142} & $\times$ & $\times$ & $\times$ \\
\SMmaterialcell{$\mathrm{UPt_{2}Si_{2}}$} & \SMidcell{0.194} & $\times$ & $\times$ & $\times$ \\
\SMmaterialcell{$\mathrm{Bi_{2}CuO_{4}}$} & \SMidcell{0.348,\allowbreak 0.694} & $\times$ & $\times$ & $\times$ \\
\SMmaterialcell{$\mathrm{UBi_{2}}$} & \SMidcell{0.378} & $\times$ & $\times$ & $\times$ \\
\SMmaterialcell{$\mathrm{UGeSe}$} & \SMidcell{0.413} & $\times$ & $\times$ & $\times$ \\
\hline
\multicolumn{5}{@{}l@{}}{\textbf{Magnetic point group }$4/m'mm$ \qquad $H=4mm$ (1 record)}\\*\hline\nopagebreak
\SMmaterialcell{$\mathrm{Co_{3}Al_{2}Si_{3}O_{12}}$} & \SMidcell{0.388} & $\times$ & $\times$ & $\checkmark$ \\
\hline
\multicolumn{5}{@{}l@{}}{\textbf{Magnetic point group }$4'/m'm'm$ \qquad $H=\bar{4}2m$ (65 records)}\\*\hline\nopagebreak
\SMmaterialcell{$\mathrm{BaMn_{2}As_{2}}$} & \SMidcell{0.18} & $\times$ & $\times$ & $\times$ \\
\SMmaterialcell{$\mathrm{CoAl_{2}O_{4}}$} & \SMidcell{0.58} & $\times$ & $\times$ & $\times$ \\
\SMmaterialcell{$\mathrm{CaMnBi_{2}}$} & \SMidcell{0.72} & $\times$ & $\times$ & $\times$ \\
\SMmaterialcell{$\mathrm{SrMnBi_{2}}$} & \SMidcell{0.73} & $\times$ & $\times$ & $\times$ \\
\SMmaterialcell{$\mathrm{U_{2}Pd_{2}In}$} & \SMidcell{0.80,\allowbreak 0.320,\allowbreak 0.625} & $\times$ & $\times$ & $\times$ \\
\SMmaterialcell{$\mathrm{U_{2}Pd_{2}Sn}$} & \SMidcell{0.81,\allowbreak 0.321} & $\times$ & $\times$ & $\times$ \\
\SMmaterialcell{$\mathrm{BaMn_{2}Bi_{2}}$} & \SMidcell{0.89} & $\times$ & $\times$ & $\times$ \\
\SMmaterialcell{$\mathrm{NpCo_{2}}$} & \SMidcell{0.126} & $\times$ & $\times$ & $\times$ \\
\SMmaterialcell{$\mathrm{Ce_{2}PdGe_{3}}$} & \SMidcell{0.166} & $\times$ & $\times$ & $\times$ \\
\SMmaterialcell{$\mathrm{CeMnAsO}$} & \SMidcell{0.186} & $\times$ & $\times$ & $\times$ \\
\SMmaterialcell{$\mathrm{GdVO_{4}}$} & \SMidcell{0.198} & $\times$ & $\times$ & $\times$ \\
\SMmaterialcell{$\mathrm{Ca_{2}MnO_{4}}$} & \SMidcell{0.211} & $\times$ & $\times$ & $\times$ \\
\SMmaterialcell{$\mathrm{Sr_{2}Mn_{3}As_{2}O_{2}}$} & \SMidcell{0.212} & $\times$ & $\times$ & $\times$ \\
\SMmaterialcell{$\mathrm{YbMnBi_{2}}$} & \SMidcell{0.267,\allowbreak 0.769} & $\times$ & $\times$ & $\times$ \\
\SMmaterialcell{$\mathrm{SrCr_{2}As_{2}}$} & \SMidcell{0.364} & $\times$ & $\times$ & $\times$ \\
\SMmaterialcell{$\mathrm{BaCr_{2}As_{2}}$} & \SMidcell{0.365} & $\times$ & $\times$ & $\times$ \\
\SMmaterialcell{$\mathrm{BaCrFeAs_{2}}$} & \SMidcell{0.366} & $\times$ & $\times$ & $\times$ \\
\SMmaterialcell{$\mathrm{EuMnBi_{2}}$} & \SMidcell{0.426,\allowbreak 2.50} & $\times$ & $\times$ & $\times$ \\
\SMmaterialcell{$\mathrm{RbFeO_{2}}$} & \SMidcell{0.456} & $\times$ & $\times$ & $\times$ \\
\SMmaterialcell{$\mathrm{CsFeO_{2}}$} & \SMidcell{0.458} & $\times$ & $\times$ & $\times$ \\
\SMmaterialcell{$\mathrm{CoRh_{2}O_{4}}$} & \SMidcell{0.461} & $\times$ & $\times$ & $\times$ \\
\SMmaterialcell{$\mathrm{MnAl_{2}O_{4}}$} & \SMidcell{0.462} & $\times$ & $\times$ & $\times$ \\
\SMmaterialcell{$\mathrm{Co_{3}O_{4}}$} & \SMidcell{0.463} & $\times$ & $\times$ & $\times$ \\
\SMmaterialcell{$\mathrm{BaMn_{2}P_{2}}$} & \SMidcell{0.464} & $\times$ & $\times$ & $\times$ \\
\SMmaterialcell{$\mathrm{HoCr_{2}Si_{2}}$} & \SMidcell{0.465,\allowbreak 0.519} & $\times$ & $\times$ & $\times$ \\
\SMmaterialcell{$\mathrm{TbPO_{4}}$} & \SMidcell{0.467} & $\times$ & $\times$ & $\times$ \\
\SMmaterialcell{$\mathrm{BaMn_{2}Sb_{2}}$} & \SMidcell{0.470} & $\times$ & $\times$ & $\times$ \\
\SMmaterialcell{$\mathrm{Ba_{2}Mn_{3}Sb_{2}O_{2}}$} & \SMidcell{0.471} & $\times$ & $\times$ & $\times$ \\
\SMmaterialcell{$\mathrm{LaMn_{2}Si_{2}}$} & \SMidcell{0.472,\allowbreak 0.498} & $\times$ & $\times$ & $\times$ \\
\SMmaterialcell{$\mathrm{EuMn_{2}Ge_{2}}$} & \SMidcell{0.474} & $\times$ & $\times$ & $\times$ \\
\SMmaterialcell{$\mathrm{ErCr_{2}Si_{2}}$} & \SMidcell{0.486} & $\times$ & $\times$ & $\times$ \\
\SMmaterialcell{$\mathrm{TbCr_{2}Si_{2}}$} & \SMidcell{0.518} & $\times$ & $\times$ & $\times$ \\
\SMmaterialcell{$\mathrm{NaCeO_{2}}$} & \SMidcell{0.525} & $\times$ & $\times$ & $\times$ \\
\SMmaterialcell{$\mathrm{CaMnSi}$} & \SMidcell{0.599,\allowbreak 0.600} & $\times$ & $\times$ & $\times$ \\
\SMmaterialcell{$\mathrm{CaMn_{2}Ge_{2}}$} & \SMidcell{0.603,\allowbreak 0.604} & $\times$ & $\times$ & $\times$ \\
\SMmaterialcell{$\mathrm{BaMn_{2}Ge_{2}}$} & \SMidcell{0.605,\allowbreak 0.606} & $\times$ & $\times$ & $\times$ \\
\SMmaterialcell{$\mathrm{BaMnSb_{2}}$} & \SMidcell{0.611} & $\times$ & $\times$ & $\times$ \\
\SMmaterialcell{$\mathrm{KMnSb}$} & \SMidcell{0.617} & $\times$ & $\times$ & $\times$ \\
\SMmaterialcell{$\mathrm{KMnBi}$} & \SMidcell{0.618} & $\times$ & $\times$ & $\times$ \\
\SMmaterialcell{$\mathrm{LaMnAsO}$} & \SMidcell{0.619,\allowbreak 0.624} & $\times$ & $\times$ & $\times$ \\
\SMmaterialcell{$\mathrm{NdMnAsO}$} & \SMidcell{0.620,\allowbreak 0.623} & $\times$ & $\times$ & $\times$ \\
\SMmaterialcell{$\mathrm{NaMnP}$} & \SMidcell{0.626,\allowbreak 0.627,\allowbreak 0.628} & $\times$ & $\times$ & $\times$ \\
\SMmaterialcell{$\mathrm{NaMnAs}$} & \SMidcell{0.629,\allowbreak 0.630} & $\times$ & $\times$ & $\times$ \\
\SMmaterialcell{$\mathrm{NaMnSb}$} & \SMidcell{0.631,\allowbreak 0.632} & $\times$ & $\times$ & $\times$ \\
\SMmaterialcell{$\mathrm{NaMnBi}$} & \SMidcell{0.634,\allowbreak 0.635} & $\times$ & $\times$ & $\times$ \\
\SMmaterialcell{$\mathrm{CeMnSbO}$} & \SMidcell{0.665} & $\times$ & $\times$ & $\times$ \\
\SMmaterialcell{$\mathrm{LaMnSbO}$} & \SMidcell{0.667} & $\times$ & $\times$ & $\times$ \\
\SMmaterialcell{$\mathrm{YbMnSb_{2}}$} & \SMidcell{0.766} & $\times$ & $\times$ & $\times$ \\
\hline
\multicolumn{5}{@{}l@{}}{\textbf{Magnetic point group }$\bar{3}'$ \qquad $H=3$ (4 records)}\\*\hline\nopagebreak
\SMmaterialcell{$\mathrm{MnTiO_{3}}$} & \SMidcell{0.19} & $\checkmark$ & $\checkmark$ & $\checkmark$ \\
\SMmaterialcell{$\mathrm{MnGeO_{3}}$} & \SMidcell{0.125} & $\checkmark$ & $\checkmark$ & $\checkmark$ \\
\SMmaterialcell{$\mathrm{MgMnO_{3}}$} & \SMidcell{0.277} & $\checkmark$ & $\checkmark$ & $\checkmark$ \\
\SMmaterialcell{$\mathrm{Yb_{3}Pt_{4}}$} & \SMidcell{0.430} & $\checkmark$ & $\checkmark$ & $\checkmark$ \\
\hline
\multicolumn{5}{@{}l@{}}{\textbf{Magnetic point group }$\bar{3}'m$ \qquad $H=3m$ (4 records)}\\*\hline\nopagebreak
\SMmaterialcell{$\mathrm{Ca_{2}YZr_{2}Fe_{3}O_{12}}$} & \SMidcell{0.751,\allowbreak 0.752} & $\times$ & $\checkmark$ & $\checkmark$ \\
\SMmaterialcell{$\mathrm{Ca_{2}LaZr_{2}Fe_{3}O_{12}}$} & \SMidcell{0.753,\allowbreak 0.754} & $\times$ & $\checkmark$ & $\checkmark$ \\
\hline
\multicolumn{5}{@{}l@{}}{\textbf{Magnetic point group }$\bar{3}'m'$ \qquad $H=32$ (9 records)}\\*\hline\nopagebreak
\SMmaterialcell{$\mathrm{Cr_{2}O_{3}}$} & \SMidcell{0.59} & $\checkmark$ & $\times$ & $\times$ \\
\SMmaterialcell{$\mathrm{Co_{4}Nb_{2}O_{9}}$} & \SMidcell{0.111} & $\checkmark$ & $\times$ & $\times$ \\
\SMmaterialcell{$\mathrm{Mn_{4}Ta_{2}O_{9}}$} & \SMidcell{0.477,\allowbreak 0.526} & $\checkmark$ & $\times$ & $\times$ \\
\SMmaterialcell{$\mathrm{U_{2}N_{2}S}$} & \SMidcell{0.484} & $\checkmark$ & $\times$ & $\times$ \\
\SMmaterialcell{$\mathrm{U_{2}N_{2}Se}$} & \SMidcell{0.485} & $\checkmark$ & $\times$ & $\times$ \\
\SMmaterialcell{$\mathrm{Mn_{4}Nb_{2}O_{9}}$} & \SMidcell{0.507} & $\checkmark$ & $\times$ & $\times$ \\
\SMmaterialcell{$\mathrm{AgRuO_{3}}$} & \SMidcell{0.733} & $\checkmark$ & $\times$ & $\times$ \\
\SMmaterialcell{$\mathrm{Na_{2}MnTeO_{6}}$} & \SMidcell{1.0.51} & $\checkmark$ & $\times$ & $\times$ \\
\hline
\multicolumn{5}{@{}l@{}}{\textbf{Magnetic point group }$6/m'$ \qquad $H=6$ (1 record)}\\*\hline\nopagebreak
\SMmaterialcell{$\mathrm{U_{14}Au_{51}}$} & \SMidcell{0.283} & $\times$ & $\times$ & $\checkmark$ \\
\hline
\multicolumn{5}{@{}l@{}}{\textbf{Magnetic point group }$6'/m$ \qquad $H=\bar{6}$ (2 records)}\\*\hline\nopagebreak
\SMmaterialcell{$\mathrm{U_{14}Au_{51}}$} & \SMidcell{0.282} & $\checkmark$ & $\checkmark$ & $\times$ \\
\SMmaterialcell{$\mathrm{K_{2}Mn_{3}(VO_{4})_{2}CO_{3}}$} & \SMidcell{1.0.21} & $\checkmark$ & $\checkmark$ & $\times$ \\
\end{longtable}
\endgroup
\FloatBarrier
\putbib[neel_josephson_diode_references]
\end{bibunit}

\end{document}